\documentclass{article}

    \usepackage[preprint,nonatbib]{neurips_2020}

\usepackage[utf8]{inputenc} 
\usepackage[T1]{fontenc}    
\DeclareUnicodeCharacter{FB02}{fl}
\DeclareUnicodeCharacter{FB01}{fi}

\usepackage[usenames,dvipsnames]{xcolor}
\usepackage{hyperref}
\definecolor{cite_color}{RGB}{0, 0, 255}
\definecolor{papergreen}{rgb}{0.886, 0.941, 0.851}
\definecolor{link_color}{RGB}{153, 0,0}  
\definecolor{url_color}{RGB}{153, 102,  0}
\definecolor{emp_color}{RGB}{0,0,255}
\hypersetup{
   colorlinks,
   citecolor=cite_color,
   linkcolor=link_color,
   urlcolor=url_color
}

\usepackage{url}            
\usepackage{booktabs}       
\usepackage{amsfonts}       
\usepackage{nicefrac}       
\usepackage{microtype}      

\usepackage{times}
\usepackage{soul}
\usepackage{mathrsfs}
\usepackage{caption}
\usepackage{graphicx}
\usepackage{amsmath}
\usepackage{amsthm}

\usepackage{colortbl}

\theoremstyle{definition}

\usepackage{booktabs}
\usepackage{algorithm}
\usepackage{algorithmic}
\usepackage{stfloats}
\usepackage{float}
\usepackage{subfigure}
\usepackage{xspace}
\usepackage{bm}
\usepackage{subfigure}
\usepackage{xcolor}
\usepackage{multirow}
\usepackage{wrapfig}

\graphicspath{{./figs/}}

\usepackage[numbers]{natbib}
\renewcommand{\citet}[1]{\cite{#1}}
\renewcommand{\citep}[1]{\cite{#1}}

\usepackage[capitalise, noabbrev]{cleveref} 
\crefname{section}{Section}{Sections}
\crefname{theorem}{Theorem}{Theorems}
\crefname{lemma}{Lemma}{Lemmas}
\crefname{equation}{Equation}{Equations}
\crefname{proposition}{Proposition}{Propositions}
\crefname{claim}{Claim}{Claims}
\crefname{assumption}{Assumption}{Assumptions}
\crefname{appendix}{Appendix}{Appendices}
\crefname{algorithm}{Algorithm}{Algorithms}
\crefname{figure}{Figure}{Figures}
\crefname{table}{Table}{Tables}
\crefname{remark}{Remark}{Remarks}
\crefname{definition}{Definition}{Definitions}
\crefname{equatinon}{Equation}{Equations}
\crefname{corollary}{Corollary}{Corollaries}

\allowdisplaybreaks

\def \m{\mathbf{m}}

\def \b{\mathbf{b}}

\def \x{\mathbf{x}}
\def \y{\mathbf{y}}

\def \e{\mathbf{e}}

\def \h{\mathbf{h}}

\def \BH{\mathbf{H}}

\def \BS{\mathbf{S}}

\def \BW{\mathbf{W}}

\def \trans{\top}

\newcommand{\pare}[1]{{(#1)}}  

\newtheorem{proposition}{Proposition}

\newcommand{\eat}[1]{}
\newcommand{\kw}[1]{{\ensuremath{\mathsf{#1}}}\xspace}

\newcommand{\DualMPNN}{\kw{MV}-\kw{GNN}}
\newcommand{\DualMPNNplus}{\kw{MV}-$\kw{GNN^{cross}}$\xspace}
\newcommand{\NodeMPN}{\kw{Node}-\kw{GNN}}
\newcommand{\EdgeMPN}{\kw{Edge}-\kw{GNN}}

\newcommand{\FLAT}{\kw{Flatten}}
\newcommand{\agt}{\text{AGG}}
\newcommand{\agg}{\text{AGG}}
\newcommand{\mlp}{\text{MLP}}

\title{Multi-View Graph Neural Networks\\ for Molecular Property Prediction}

\author{%
  Hehuan Ma\\
  University of Texas at Arlington\\
  Arlington, TX 76019 \\
  \texttt{hehuan.ma@mavs.uta.edu} \\
    \And
  Yatao Bian \thanks{These authors contributed equally to this work.}\\
  Tencent AI Lab\\
  Shenzhen, China 518057 \\
  \texttt{yatao.bian@gmail.com} \\
    \And
  Yu Rong \footnotemark[1]\\
  Tencent AI Lab\\
  Shenzhen, China 518057 \\
  \texttt{yu.rong@hotmail.com} \\
    \And
 Wenbing Huang \\
  Department of Computer Science and Technology\\
  Tsinghua University\\
  Beijing, China \\
  \texttt{hwenbing@126.com} \\
    \And
 Tingyang Xu \\
  Tencent AI Lab\\
  Shenzhen, China 518057 \\
  \texttt{tingyangxu@tencent.com} \\
    \And
 Weiyang Xie \\
  Tencent AI Lab\\
  Shenzhen, China 518057 \\
  \texttt{weiyangxie@tencent.com} \\
    \And
 Geyan Ye \\
  Tencent AI Lab\\
  Shenzhen, China 518057 \\
  \texttt{blazerye@tencent.com} \\
    \And
  Junzhou Huang \thanks{Corresponding author.}\\
  University of Texas at Arlington\\
  Arlington, TX 76019 \\
  \texttt{jzhuang@uta.edu} \\
}
\begin{document}

\maketitle
\setcounter{footnote}{0}

\begin{abstract}
 The crux of molecular property prediction is to generate meaningful representations of the molecules. One promising route is to exploit the molecular graph structure through  Graph Neural Networks (GNNs).
 It is well known that both atoms and bonds significantly affect the chemical properties of a molecule, so an expressive model shall be able to exploit both node (atom) and edge (bond) information simultaneously.
 Guided by this observation, we present \textbf{M}ulti-\textbf{V}iew \textbf{G}raph \textbf{N}eural \textbf{N}etwork (\DualMPNN), a multi-view message passing architecture to enable more accurate predictions of molecular properties. In \DualMPNN, we introduce a shared self-attentive readout component and disagreement loss to stabilize the training process. This readout component also renders the whole architecture interpretable. We further boost the expressive power of \DualMPNN by proposing a \emph{cross-dependent message passing scheme} that enhances information communication of the two views, which results in the \DualMPNNplus variant. 
 Lastly, we theoretically justify the expressiveness of the two proposed models in terms of distinguishing non-isomorphism graphs. 
 Extensive experiments demonstrate that \DualMPNN models achieve remarkably superior performance over the state-of-the-art models on a variety of challenging benchmarks. Meanwhile, visualization results of the node importance are consistent with prior knowledge, which confirms the interpretability power of \DualMPNN models.
\end{abstract}

\vspace{-0.51cm}
\section{Introduction}

Molecular property prediction is a  challenging task in  drug discovery, and attracts increasingly more attention in the last decades.
For example, designing molecular fingerprints based on the radial group of the molecular structure, then use the converted fingerprint for property prediction \citep{glen2006circular}. Specifically, a particular property of a given molecule is identified by applying specific models. However, traditional molecular property prediction methods usually i) requires chemical experts to conduct professional experiments to validate the property label, ii) desires high R\&D cost and massive amount of time, and iii) asks for specialized model for different properties, which lacks generalization capacity \citep{paul2010improve, ching2018opportunities}. 

To date, Graph Neural Networks (GNNs) have gained increasingly more popularity due to its capability of modeling graph structured data. Successes have been achieved in various domains, such as social network \citep{ velivckovic2017graph,NIPS2018_7707}, knowledge-graphs \citep{guu2015traversing, hamilton2018embedding}, and recommendation systems \citep{mao2016multirelational, monti2017geometric}. Molecular property prediction is also a promising application of GNNs since a molecule could be represented as a topological graph  by treating atoms as nodes, and bonds as edges. Compared with other representations for molecules, such as SMILES \citep{neglur2005assigning}, which represents molecules as sequences but losses structural information, graph representation of molecules can naturally capture the information from the molecular structure, including both the nodes (atoms) and edges (bonds). In this sense, a molecular property prediction task is equivalent to a supervised graph classification problem (see, for example, toxicity prediction \citep{pires2015pkcsm} and protein interface prediction \citep{fout2017protein}).

\begin{wrapfigure}{r}{0.4\textwidth}
\centering
\includegraphics[width=0.4\textwidth]{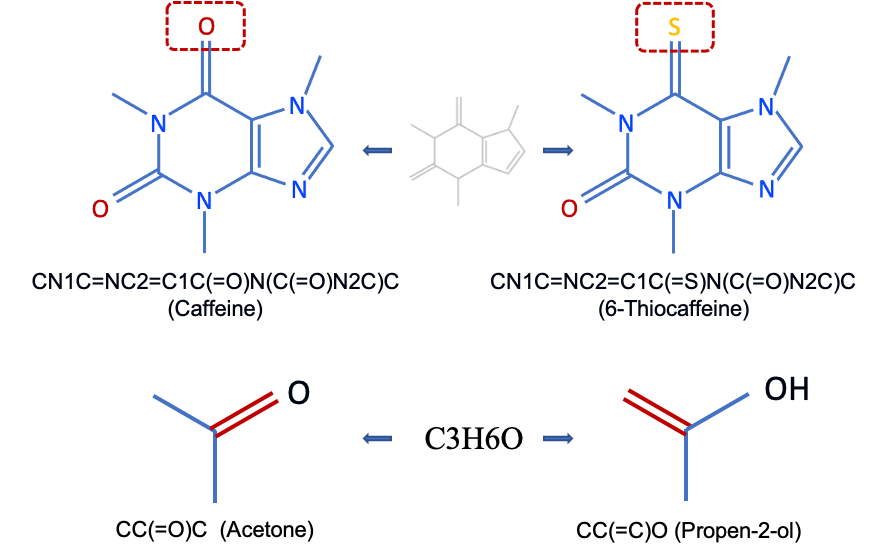} 
\vspace{-1ex}
\caption{\footnotesize The upper two molecules share same bond structures, but contains different atoms. The lower two molecules share same atoms, but equip with different bonds.}
\vspace{-1ex}
\label{compares}
\end{wrapfigure}

Despite the fruitful results obtained by GNNs, there remains two limitations: 1) Most of the GNN models focus either on the embedding of nodes or edges. However, in many practical scenarios, nodes and edges play equally important roles. For example, in a knowledge graph, a node represents an entity, and the edge indicates the interact ontologies and semantics between linked nodes. Different edges that represent different relations hence may lead to different answers. Especially, molecular property prediction also demands information from both atoms and bonds to generate precise graph embeddings. 
Molecules with different atoms (nodes) but same bonds (edges) are distinct compounds with different properties and so as to different bonds (edges) but same atoms (nodes). As shown in \cref{compares}(upper), equipped with same bonds, only one-atom difference make the two molecules distinct Octanol/Water Partition Coefficients. Caffine is more hydrophilic while 6-Thiocaffeine is more lipophilic \citep{bhal2007logp}. Similarly,  in \cref{compares}(lower), the molecular formulas of Acetone and Propen-2-ol are exactly the same, but the bond difference makes Acetone behave mild irritation to human eyes, nose, skin, etc. Accordingly, both nodes and edges are fairly essential for molecular property prediction. Therefore, how to properly integrate \emph{both node and edge information in a unified manner} is the first challenge. 2) Exsiting GNNs usually lack interpretability power, which is actually crucial for drug discovery tasks. 
Take molecular property prediction as an example, being aware of how the model validate the property will help practitioners figure out the key components that determine certain properties \citep{preuer2019interpretable}.

In pursuit of tackling the above challenges, we propose a new multi-view architecture: \DualMPNN, which considers the diversity of different aspects for one single target \citep{sun2013survey}.
\DualMPNN consists of two sub-modules that generate the graph embeddings from node and edge, respectively. Therefore, it investigates the molecular graph from two views simultaneously. Meanwhile, we design a shared self-attentive readout component to produce the graph-level embedding and   interpretability results as well. To stabilize the training process of the multi-view architecture, we present a disagreement loss to restrain the difference of the predictions between two sub-modules. 
Furthermore, we propose a cross-dependent message passing scheme to enable more efficient information communication between different views, the resulted variant  is termed as \DualMPNNplus. 
Comprehensive experiments on 11 benchmarks demonstrate the superiority of \DualMPNN and \DualMPNNplus. 

Overall, our main contributions are: 
    1) We propose \DualMPNN, a multi-view architecture for molecular property predictions. It involves a shared self-attentive readout component that produces interpretable results, and a disagreement loss to stabilize the training process of the two-view pipeline.
    2) In order to encourage information communication in \DualMPNN, we propose a cross-dependent message passing scheme, which constitutes the variant \DualMPNNplus. It is empirically demonstrated to have superior expressive power than \DualMPNN.
    3) In terms of theories on expressive power, we show that \DualMPNN is at least as powerful as the well-justified Graph Isomorphism Network (GIN) \citep{xu2018powerful}, and \DualMPNNplus is strictly more powerful than GIN. 
    4) Extensive experiments on 11 benchmark datasets validate the effectiveness of \DualMPNN models. Namely, the overall performance of \DualMPNN and \DualMPNNplus achieve up to 3.6\% improvement on classification benchmarks and 28.7\%  improvement on regression benchmarks compared with  SOTA methods. Moreover, case studies on toxicity prediction demonstrate the interpretability power of \DualMPNN and \DualMPNNplus.

\section{Preliminaries on Molecular Representations and Generalized GNNs} \label{preliminary}

We abstract a molecule $c$  as a topological graph  $G_c=(\mathcal{V}, \mathcal{E})$, where $|\mathcal{V}|=p$ refers to the set of $p$ nodes (atoms) and $|\mathcal{E}|=q$ refers to a set of $q$ edges (bonds). 
 $\mathcal{N}_v$ denotes the neighborhood set of node $v$. 
We denote the feature of node $v$ as ${\x}_v \in \mathbb{R}^{d_n}$ and the feature of edge $(v, k)$ as ${\e}_{vk}\in \mathbb{R}^{d_e}$\footnote{With a bit abuse of notations, ${\e}_{vk}$ can represent either the edge $(v, k)$ or the edge features.}. $d_n$ and $d_e$ refer to the feature dimensions of nodes and edges, respectively. 
Exemplar node and edge features are the chemical relevant features such as atomic mass and bond type. Please refer to \cref{feature_extraction} for detailed feature extraction process. 
Properties of a molecule ${\y}$ constitute the targets of the predictive task. 
Given a molecule $c$ and its associated graph representation $G_c$, molecular property prediction aims to predict the properties ${\y}_c$ according to the  embedding $\xi_c$ of $G_c$.
The values of ${\y}$ are either categorical values (e.g., toxicity and permeability \citep{richard2016toxcast,martins2012bayesian}) for classification tasks or real values (e.g., atomization energy and the electronic spectra \citep{blum,ramakrishnan2015electronic}) for regression tasks.

\textbf{Generalized GNNs.}
%
Most of the GNN models are built upon the message passing process, 
which aggregates and passes the feature information of corresponding neighboring nodes to produce new hidden states of the nodes. 
After the message passing process,  all hidden states of the nodes are fed into a readout component, to produce the final graph-level embedding.
Here we present a generalized version of the message passing scheme.
Suppose there are $L$ iterations/layers, and iteration $l$ contains $K_l$ hops.
In iteration $l$, the $k$-th hop of message passing can be formulated as,
\begin{align}\label{eq_message_passing}
  \text{(Message Aggregation)} \quad   & \m_v^{\pare{l, k}} = \agt^\pare{l}( \{ \h_v^\pare{l, k-1}, \h_u^\pare{l, k-1}, \e_{uv} \;|\;  u\in \mathcal{N}_v  \}  ),
  \\   \text{(State Update)} \quad   & \h_v^\pare{l, k} = \text{MLP}^\pare{l}(\m_v^{\pare{l, k}}),
\end{align}
where we make the convention that $\h_v^\pare{l, 0} := \h_v^\pare{l-1, K_{l-1}}$. $\agg^\pare{l}$ denotes the aggregation function, 
$\m_v^{\pare{l, k}}$ is the aggregated message, and $\mlp^\pare{l}$ is a multi-layer perceptron\footnote{For instance, it could be a one layer neural net, then the state update becomes $\h_v^\pare{l, k} = \sigma( \BW^\pare{l}\m_v^{\pare{l, k}} + \b^\pare{l})$, where $\sigma$ stands for the activation function.}.   
There are several popular choices for the  aggregation function $\agg^\pare{l}$, such as mean, max pooling and the graph attention mechanism \citep{velivckovic2017graph}. 
Note that 
for one iteration of message passing, there are a layer of trainable parameters (parameters inside $\agg^\pare{l}$ and $\text{MLP}^\pare{l}$). These parameters are shared across the $K_l$ hops within iteration $l$. 
After $L$ iterations of message passing, the hidden states of the last hop in the last iteration are used as the embeddings of the nodes, i.e., $\h_v^\pare{L, K_L}, v\in \mathcal{V}$. Lastly, a READOUT operation is applied to generate the graph level representation,
\begin{align}\label{eq_readout}
    \h_G = \text{READOUT}( \{ \h_v^\pare{0, K_0}, ..., \h_v^\pare{L, K_L}  \;|\; v \in \mathcal{V}  \} ).
\end{align}
If choosing the sum aggregation with a learnable parameter $\epsilon^\pare{l}$, i.e., $\agt^\pare{l}( \{ \h_v^\pare{l, k-1}, \h_u^\pare{l, k-1}, \e_{uv} |  u\in \mathcal{N}_v  \}) =  ( (1 + \epsilon^\pare{l}) \h_v^\pare{l, k-1} + \sum_{u\in \mathcal{N}_v} \h_u^\pare{l, k-1}) || (\sum_{u\in \mathcal{N}_v} \e_{uv})$ ($||$ is the concatenation operation),  
then generalized GNN recovers graph isomorphism network (GIN) architecture \citep{xu2018powerful}, which provably generalizes the WL graph isomorphism test \citep{weisfeiler1968reduction}. 

\section{Multi-View GNN (\DualMPNN) and its Variant \DualMPNNplus}
\label{sec_technical}


In this section, we will first introduce the high-level framework of \DualMPNN models,  then illustrate each of its components in detail. Lastly we theoretically  verify their expressive power.

\begin{figure}[!t]
\vspace{-2em}
\centering
\includegraphics[width=0.9\textwidth]{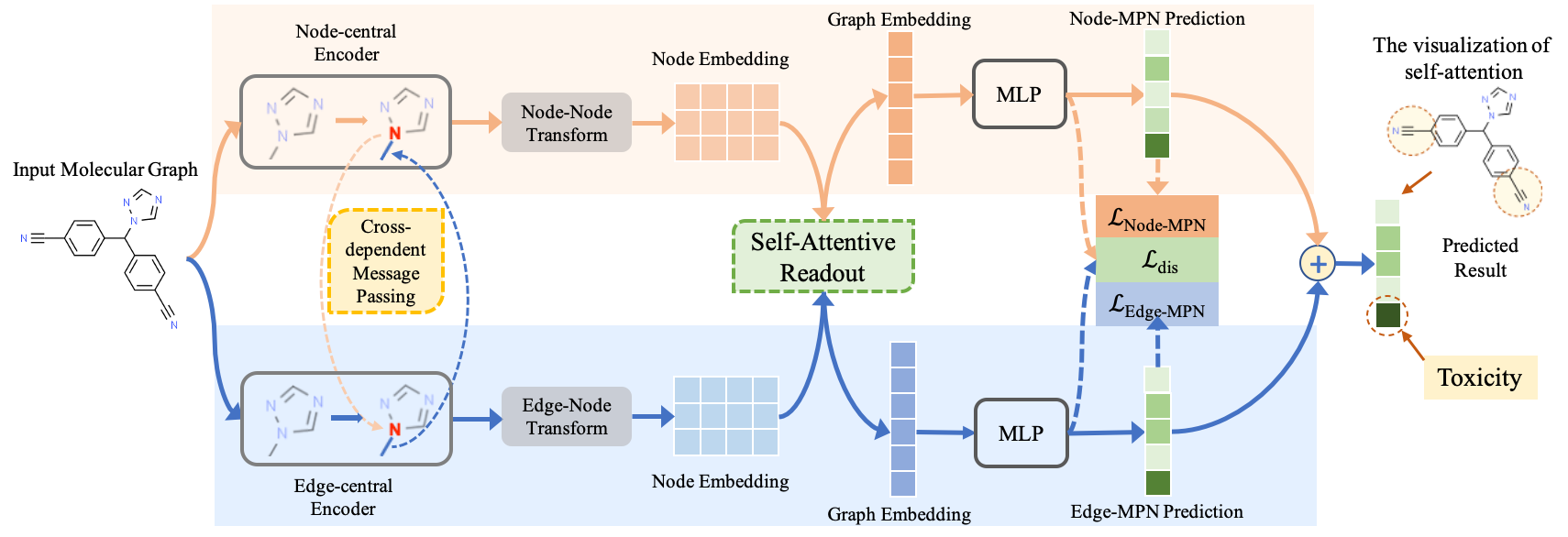} 
\caption{\footnotesize
Overview of \DualMPNN models.
\DualMPNN models pass the graph through two encoders to generate two sets of node embeddings. A shared self-attention readout learns the node importance and produce two graph embeddings accordingly. The embeddings are then fed into two MLPs to make predictions. The final prediction is the ensemble of the two predictions. 
Furthermore, by visualizing the learned attentions over nodes, one can identify the atoms/functional groups that are responsible for the predictions. For example, \DualMPNN finds out that the cyano groups contribute to the toxicity significantly.}
\label{fig.model}
\vspace{-1em}
\end{figure} 

\subsection{Overview of the Multi-View Architecture}


The multi-view architecture equally  considers both atom features 
and bond features for constituting a molecular representation.
As shown in \cref{fig.model}, the proposed  architecture contains two concurrent sub-modules, \emph{Node-central encoder} and \emph{Edge-central encoder}, which output the node/edge embedding matrix from the graph topology as well as node/edge features. 
Next, \DualMPNN adopts an aggregation function to produce the graph embedding vector from the node/edge embedding matrix. Other than the mean-pooling mechanism, we propose to use the \emph{self-attentive aggregation} to learn different weights of the node/edge embeddings to produce the final graph embedding. Furthermore, the self-attentive aggregation layer is shared between the 
node-central and edge-central encoders, to reinforce the learning of the node features and the edge features, respectively.
After the self-attentive aggregation, \DualMPNN feeds the graph embedding from the node-central encoder and the edge-central encoder to two MLPs to fit the loss function. To stabilize the training process of this multi-view architecture, we employ the \emph{disagreement loss} to enforce the outputs of the two MLPs to be close with each other.

\subsection{Node-central and Edge-central Encoders}\label{sec.encoders}

To ease the exposition, in the sequel when using one singe superscript we mean the hop index $k$ while ignoring the layer/iteration index $l$. 

\textbf{{Node-central Encoder.}} 
\NodeMPN is built upon the generalized 
message passing in \cref{eq_message_passing}. Additionally, we add \emph{input} and \emph{output} layers, to enhance its expressive power.
Specifically, 
\begin{align}\label{eq_node_mpnn}
 \m_v^{\pare{k}} = \text{\agt}_{\text{node}}(\{ \h_v^\pare{k-1}, \h_u^\pare{k-1}, \e_{uv} \;|\;  u\in \mathcal{N}_v \}  ), \quad  \h_v^\pare{k} = \mlp_{\text{node}}( \{\m_v^{\pare{k}},  \h_v^\pare{0}\}), 
\end{align}
where ${\h}_v^{(0)}=\sigma({\BW}_{\text{nin}}{\x}_v)$ is the input state of \NodeMPN, ${\BW}_{\text{nin}}\in \mathbb{R}^{d_{\text{hid}}\times d_n}$ is the input weight matrix. The input layer can also be viewed as a residual connection.
%
After $L$ iterations of message passing, we utilize an additional message passing step with a new weight matrix ${\BW}_{\text{nout}}\in \mathbb{R}^{d_{\text{out}} \times (d_n+d_{\text{hid}})}$ to produce the final node embeddings:
\begin{align}
    {\m}_{v}^{\text{o}} &= \agt_{\text{node}}( \{ {\h}_{u}^{(L, K_L)}, {\x}_u | u\in \mathcal{N}_v \}),\quad
    {\h}_v^{\text{o}} = \sigma({\BW}_{\text{nout}} {\m}_v^{\text{o}}).
\end{align}
We denote ${\BH}_{n}=[{\h}_1^{\text{o}},\cdots, {\h}_p^{\text{o}}] \in \mathbb{R}^{d_{\text{out}} \times p}$ as the output embeddings of \NodeMPN, where $d_{\text{out}}$ is the dimension of output embeddings.   



\textbf{{Edge-central Encoder.}}
 In classical graph theory, the line graph $L(G)$ of a graph $G$ is the graph that encodes the adjacencies between edges of $G$ \citep{Harary1960}. $L(G)$ provides a fresh perspective to understand the original graph, i.e., the nodes are  viewed as the connections while edges are viewed as entities. Therefore, it enables to perform  message passing operation through edges to imitate \NodeMPN on $L(G)$ \citep{yang2019analyzing}. Namely, given an edge $(v,w)$, we can formulate the Edge-based GNN (\EdgeMPN) as:
 \begin{align}
     {\m}_{vw}^{(k)} &= \agt_{\text{edge}}(\{{\h}_{vw}^{(k-1)}, {\h}_{uv}^{(k-1)}, {\x}_u | u \in \mathcal{N}_v \setminus w\}),\quad
     {\h}_{vw}^{(k)} = \mlp_{\text{edge}}(\{ {\m}_{vw}^{(k-1)}, {\h}_{vw}^{(0)} \}),
     \label{equ_bondmpnn}
 \end{align}
 where ${\h}_{vw}^{(0)}=\sigma({\BW}_{\text{ein}}{\e}_{vw})$ is the input state of \EdgeMPN,  ${\BW}_{\text{ein}}\in \mathbb{R}^{d_{\text{hid}}\times d_e}$ is the input weight matrix.  
 In~\cref{equ_bondmpnn}, the state vector is defined on edge $\e_{vw}$ and the neighboring edge set of $\e_{vw}$ is defined as all edges connected to the start node $v$ except the node $w$. 
 \cref{fig:dmpnn} in \cref{app_details_models}  shows an example of the message passing process in \EdgeMPN. 
 
 After recurring $L$ steps of message passing, the output of \EdgeMPN is the state vectors for edges. In order to incorporate the shared-attentive readout to generate the graph embedding, one more round of message passing on nodes is employed to transform edge-wise embeddings to node-wise embeddings, and generate the second set of node embeddings. Specifically,
 \begin{align}\label{eq_edge_node_transform}
  \text{(Edge-Node transform)} \quad  {\m}_v^\text{o} &= \agt_{\text{edge}}({\h}_{uv}^{(L, K_L)}, {\x}_u | u \in \mathcal{N}_v ), \quad
     {\h}_v^\text{o} = \sigma({\BW}_{\text{eout}}{\m}_{v}^\text{o}),
 \end{align}
 where ${\BW}_{\text{eout}}\in \mathbb{R}^{d_{\text{out}}\times(d_n + d_\text{hid})}$ specifies the weight matrix. Therefore, the final output of \EdgeMPN provides a new set of \emph{node} embeddings from the edge message passing process. This set of node embeddings are denoted as ${\BH}_{e}=[{\h}_1^{\text{o}},\cdots, {\h}_p^{\text{o}}] \in \mathbb{R}^{d_{\text{out}} \times p}$.

\eat{
\textbf{Bond Message Passing.}
Similar with atom messages, we can define the bond message passing model $\mathscr{F}_{b}^D(G)$ to model the bond features. Namely, given a bond $e_{vw}$, the message $m^{d}_{vw}$ and state $h^{d}_{vw}$ can be modeled as:
\begin{align}
    \notag h_{vw}^{0} &= \sigma(W_{in}e_{vw})\\
    \notag m_{vw}^{d+1} &= \sum_{k \in N(v) \setminus w } \text{cat}(h_{kv}^{d}, x_k),\\
    h_{vw}^{d+1} &= \sigma(W_{b}m_{vw}^{d+1}+h_{vw}^{0}).
    \label{equ:bondmpnn}
\end{align}
In~\eqref{equ:bondmpnn}, we define the neighbor bond set of $e_{vw}$ by all bonds connected to the start atom $v$ except $w$.\footnote{For the directed graph, we only count bond ended with atom $v$.} The attached feature $\mu_\text{attached}$ is the atom feature $x_{k}$. The message updating function and state updating function is similar with atom message passing.

After recurring D steps for the message passing phase, one more round aggregation on atoms is employed to transform bond-wise message to atom-wise message, and generate the final atom embeddings $h_v^{o}$:
\begin{align}
    \notag m_v^{o} &= \sum_{k \in N(v)} \text{cat}(h_{ki}^D, x_k)\\
    \notag h_v^{o} &= \sigma(W_om_v^{o}),
\end{align}
where $W_o$ is the output parameter, and the final output of $\mathscr{F}_{b}^D(G)$ is represented as $H_b=[h_1^{o},\cdots, h_n^{o}]$. 
}

\subsection{Interpretable Readout Component for Generating Graph-level Embedding}


To obtain a fixed length graph representation, a readout component is usually employed on the node embeddings. 
In this work, we considered two readout transformations to obtain the molecular representation. 
The first is the simple \emph{mean-pooling readout},  
the molecular representation is given by $  {\xi}_n = \frac{1}{p}\sum_{{\h}_i^o\in {\BH}_n} {\h}_i^\text{o}$.
However, the average operation tends to produce smooth outputs. Therefore, it diminishes the expressive power. 
To overcome the drawbacks of mean-pooling, we develop the   \emph{interpretable self-attentive readout component} based on the attention mechanism \citep{velivckovic2017graph,li2019semi}. Namely, given an output of node-central encoder ${\BH}_n \in \mathbb{R}^{d_{\text{out}} \times p}$, the self-attention ${\BS}$ over nodes is: 
\begin{align}
    {\BS}=\operatorname{softmax}\left({\BW}_{2} \tanh \left({\BW}_{1} {\BH}_n\right)\right),
    \label{sa}
\end{align}
where ${\BW}_{1} \in \mathbb{R}^{d_{\text{attn}} \times d_{\text{out}}}$ and ${\BW}_{2} \in \mathbb{R}^{r \times d_{\text{attn}}}$ are learnable matrices. In~\cref{sa}, ${\BW}_{1}$  linearly transforms the node embeddings from $d_{\text{out}}$-dimensional space to a $h_{\text{attn}}$-dimensional space. 
${\BW}_{2}$ provides $r$ different insights of node importance, then followed by a softmax function to normalize the importance.
To enable the feature information extracted from node and edge encoders communicating during the multi-view training process, we \emph{share} the parameters ${\BW}_{1}$ and ${\BW}_{2}$ between the two sub-models.
Given ${\BS}$, we can obtain the graph-level embedding
by $\bm{\xi}_n = \FLAT({\BS}{\BH}_n^\trans)$. 
The self-attention ${\BS}$ implies importance of the nodes when generating graph embedding,  hence  indicating contributions of the nodes for downstream tasks, which equips \DualMPNN with interpretability power.

\eat{
Different from mean-pooling, we employ the self-attention mechanism to learn the atom importance and encode atom embedding into a size-invariant molecular representation. Given a output of node-MPN model $H_a \in \mathbb{R}^{n \times a}$, the self-attention $S_a$ is defined as:
\begin{align}
    S_a=\operatorname{softmax}\left(W_{2} \tanh \left(W_{1} H_a^{T}\right)\right),
    \label{sa}
\end{align}
where $W_{1} \in \mathbb{R}^{h_{\text{attn}} \times a}$ and $W_{2} \in \mathbb{R}^{r \times h_{\text{attn}}}$ are two parameters. In~\eqref{sa}, $W_{1}$ is to linearly transform the atom embedding from $a$-dimensional space to a $h_{\text{attn}}$-dimensional space.  $tanh$ function is equipped for introducing nonlinearity. $W_{2}$ brings $r$ different insights of atom importance, then followed a softmax function to normalize the importance, which makes the summation of different importance views to 1. To enable the feature information extracted from DMPNN and MPNN binding and communicating during the dual training process, we share the attention parameters between two sub-models.

According to the self-attention matrix $S_a$, we can obtain the \textbf{size invariant} and \textbf{atom importance involved} molecule graph representation $\xi_a \in \mathbb{R}^{r\times a}$ by 
\begin{align}
    \xi_a = \text{flatten}(S_aH_a)
\end{align}
}

\subsection{The Disagreement Loss for \DualMPNN Models}

Suppose the dataset contains graphs $\mathcal{G}=\{G_i\}_{i=1}^{K}$ and corresponding labels $\mathcal{Y} = \{{\y}_i\}_{i=1}^{K}$. 
Given one graph $G_i$, due to the nature of the multi-view architecture, we obtain two graph embeddings $\bm{\xi}_n$ and $\bm{\xi}_e$, from 
the node message passing and edge message passing, respectively. Feeding them into the MLPs results in two predictions $\bm{\gamma}_{n,i}$ and $\bm{\gamma}_{e,i}$ for the same target $\y_i$. Naturally, the losses should get the supervised prediction loss involved, i.e., 
$\mathcal{L}_{\text{pred}} = \sum_{G_i \in \mathcal{G}}(\mathcal{L}_{\kw{Node}\text{-}\kw{GNN}}({\y}_i, \bm{\gamma}_{n, i}) +\mathcal{L}_{\kw{Edge}\text{-}\kw{GNN}}({\y}_i, \bm{\gamma}_{e, i}) )$. The specific loss function $\mathcal{L}_{\kw{Node}\text{-}\kw{GNN}}$ and $\mathcal{L}_{\kw{Edge}\text{-}\kw{GNN}}$ should depend on the task types, say, cross-entropy for classification and mean squared error for regression.

However, with only the $\mathcal{L}_{\text{pred}}$ loss, 
we observed  unstable behaviors of the training process, which is caused by the loose 
constraint of the node and edge message passings. 
To resolve this problem, we propose the
\emph{disagreement loss}, which is responsible for restraining the two predictions from node-central and edge-central encoders. 
Specifically, we employ the mean squared error $\mathcal{L}_{\text{dis}}=\sum_{G_i \in \mathcal{G}} \left \vert  \gamma_{n, i}- \gamma_{e, i}\right \vert ^{2}$.
%
Overall, the shared self-attentive readout and the disagreement loss alleviate the node variant dependency, and reinforce the restriction during the training process to promise the model converge to a stationary status. 
Finally, the overall loss function contains two parts: $\mathcal{L}=\mathcal{L}_{\text{pred}} + \lambda \mathcal{L}_{\text{dis}}$, 
where 
$\lambda$ is a tradeoff hyper-parameter.

\subsection{\DualMPNNplus: \DualMPNN Equipped with the Cross-dependent Message Passing Scheme}

Though \DualMPNN is proved to have superior performance for many molecular property prediction tasks (as verified in the experiments), we find that 
the information flow in \DualMPNN is not sufficiently efficient.
%
Suppose all the information needed to predict the property resides in the molecule itself. For \DualMPNN, the
information flows through two distinct paths in parallel: one path is the \NodeMPN encoder, the other one is the \EdgeMPN encoder. The information from the two paths finally joins at the disagreement loss. 

However, the two flows of information could meet \emph{earlier}, to 
enable more efficient information communication.
The strategy to implement this is the \emph{cross-dependent message passing} scheme. On a high level, it makes the message passing operations of the node and edge cross-dependent with each other. Specifically, we change the message passing operations of the node and edge encoders (in Equations \eqref{eq_node_mpnn} and \eqref{equ_bondmpnn}, respectively) to be:
\begin{align}\label{equ_cross_dependent}
     {\m}_{v}^{(k)} &=  \agt_{\text{node}}(\{{\h}_{v}^{(k-1)}, {\h}_{u}^{(k-1)}, \textcolor{blue}{{\h}_{vu}^{(k-1)}}, {\e}_{vu} | u \in \mathcal{N}_v \}),  \;
    \h_v^\pare{k} = \mlp_{\text{node}}( \{\m_v^{\pare{k}},  \h_v^\pare{0}\}),\\\notag 
 {\m}_{vw}^{(k)} &= \agt_{\text{edge}}(\{{\h}_{vw}^{(k-1)}, {\h}_{uv}^{(k-1)}, \textcolor{blue} {{\h}_{u}^{(k-1)}}, {\x}_u | u \in \mathcal{N}_v \setminus w \}), \; {\h}_{vw}^{(k)} = \mlp_{\text{edge}}(\{ {\m}_{vw}^{(k)}, {\h}_{vw}^{(0)} \}).
\end{align}
The first row indicates new node message passing, while the second row
shows edge message passing. One can see that when applying aggregation in node message passing, we use the newest hidden states of edges (blue colored). While conducting aggregation in edge message passing, it requires the newest hidden states of nodes. In this way, the two paths of information flow become cross-dependent with each other. 
We will empirically show that the cross-dependent message passing scheme enables more expressive power compared to the vanilla \DualMPNN architecture.

\subsection{Expressive Power of \DualMPNN and \DualMPNNplus}
\label{sec_capacity}

\DualMPNN and \DualMPNNplus achieve superior performance 
compared to all baselines in the experiments. 
In this section, 
we justify its performance by studying their expressiveness under the framework of distinguishing non-isomorphic 
graphs. By comparing the expressive power with the 
well justified architecture GIN \citep{xu2018powerful}, 
we reach the following conclusions. 
\begin{proposition}\label{prop_expressiveness}
 In terms of expressive power of models, the following conclusions hold:
 \begin{enumerate}
     \item \DualMPNN is at least as powerful as  the Graph Isomorphism Network (GIN of \cite{xu2018powerful}), which provably generalizes the WL graph isomorphism test.

     \item  \DualMPNNplus  is \emph{strictly} more powerful than the Graph Isomorphism Network.
 \end{enumerate}
\end{proposition}
Detailed proof is deferred to \cref{appendix_proof}. \cref{prop_expressiveness} shows that \DualMPNN models
have sufficient model capacity in terms of distinguishing graphs  compared to the GIN architecture and the WL graph isomorphism test.
This observation also explains why it reaches superior performance on various baseline tasks, which will be further verified in the experiments. 

\section{Experimental Results}
\label{sec_experiments}

We conduct the performance evaluations of \DualMPNN and \DualMPNNplus with various SOTA baselines on molecular property classification and regression tasks. Due to the space limitation, the results of the regression tasks are deferred to \cref{additional_regression}. We also preform the ablation studies on different components of the \DualMPNN models. Lastly, we conduct case studies to demonstrate the interpretability power of the proposed models. 
Source code will be released soon. 


\textbf{Datasets.}
We experimented with 11 popular benchmark datasets, among which six are classification tasks and the others are regression tasks. Specifically, \texttt{BACE} is about the biophysics property; \texttt{BBBP}, \texttt{Tox21}, \texttt{Toxcast}, \texttt{SIDER}, and \texttt{Clintox} record several molecular physiology properties; \texttt{QM7} and \texttt{QM8} contain molecular quantum mechanics information; \texttt{ESOL}, \texttt{Lipophilicity} and \texttt{Freesolv} document physical chemistry properties \citep{wu2018moleculenet}. Details are deferred to \cref{dataset_description}.

\textbf{Baselines.}
We thoroughly evaluate the performance of our methods against popular baselines from both machine learning and chemistry communities. Among them, Inﬂuence Relevance Voting (\kw{IRV}) \citep{swamidass2009influence}, \kw{LogReg} \citep{friedman2000additive}, Random Forest (\kw{RF}/\kw{RF\_Reg}) \citep{breiman2001random} utilize different traditional machine learning approaches. \kw{GraphConv} \citep{duvenaud2015convolutional}, \kw{Weave} \citep{kearnes2016molecular}, \kw{SchNet} \citep{schutt2017schnet}, \kw{MGCN} \citep{lu2019molecular}, \kw{N}-\kw{Gram} \citep{liu2019n}, \kw{MPNN} \citep{gilmer2017neural} and \kw{DMPNN} \citep{yang2019analyzing} are GNN-based models. For \kw{SchNet} and \kw{MGCN}, we use DGL \citep{wang2019dgl} implementations; for \kw{N}-\kw{Gram} and \kw{DMPNN}, we use open source codes provided by the author; for \kw{MPNN}, we use the implementation by \citep{yang2019analyzing}; for others, we use the MoleculeNet \citep{wu2018moleculenet} implementations. Details can be found in \cref{baseline_model}. 


\textbf{Dataset Splitting.} We apply the scaffold splitting for all tasks on all datasets, which is more practical and challenging than 
random splitting. More details about this splitting method is introduced in \cref{dataset_description}.
%
\textbf{Evaluation Metrics.}
All classification task are evaluated by AUC-ROC. For the regression task, we apply MAE and RMSE to evaluate the performance of regression task on different datasets.

\subsection{Performance Evaluation on Classification Tasks}

To demonstrate the effectiveness of shared self-attentive readout and the disagreement loss, we also implement two naive schemes. \kw{Concat+Mean} concatenates the mean-pooling outputs of the two sub-modules, and \kw{Concat+Attn} concatenates the self-attentive outputs\footnote{We do not share the attention here.} of the two sub-modules. \cref{baseline} summarizes the results of the classification tasks. To evaluate the robustness of our method, we report the mean and standard deviation of 10 times runs with different random seeds for \DualMPNN, \DualMPNNplus and the variants. \cref{baseline} implies  the following observations: (1) our \DualMPNN models gain significant enhancement against SOTAs on all datasets consistently, \DualMPNNplus even performs slightly better than \DualMPNN. Specifically, \DualMPNN gains the average AUC boost by $1.15\%$ on average compared with the SOTAs on each dataset, while \DualMPNNplus improves it to $1.65\%$, which is regarded as the remarkable boost, considering the challenges on these benchmarks.  (2) Compared with the SOTAs, \DualMPNN and \DualMPNNplus has much smaller standard deviation, which implies that our models are more robust than the baselines. (3) Compared with the two simple variants, \DualMPNN and \DualMPNNplus demonstrate the superiority both on performance and robustness. It validates the effectiveness of the multi-view architecture with disagreement loss constraint. 


\begin{table}[!h]
\centering
\captionsetup{type=table}
\renewcommand\arraystretch{1.2}
\caption{\footnotesize Performance of classification tasks on AUC-ROC (higher is better) with the scaffold split. Best score is marked as \textbf{bold}, and the best baseline is marked in gray. Green cells indicate the results of our methods.}
\vspace{1.5ex}
\resizebox{0.8\textwidth}{!}{ 
\begin{tabular}{ccccccc}
\toprule
Method & \texttt{BACE} & \texttt{BBBP} & \texttt{Tox21} & \texttt{ToxCast} & \texttt{SIDER} & \texttt{ClinTox} \\ 
\midrule
\kw{IRV} & 0.838$_{\pm0.055}$ & 0.877$_{\pm0.051}$ & 0.699$_{\pm0.055}$ & 0.604$_{\pm0.037}$ & 0.595$_{\pm0.022}$ & 0.741$_{\pm0.069}$ \\
\kw{LogReg} & 0.844$_{\pm0.040}$ & 0.835$_{\pm0.067}$ & 0.702$_{\pm0.028}$ & 0.613$_{\pm0.033}$ & 0.583$_{\pm0.034}$ & 0.733$_{\pm0.084}$ \\
RF & 0.856$_{\pm0.019}$ & 0.881$_{\pm0.050}$ & 0.744$_{\pm0.051}$ & 0.582$_{\pm0.049}$ & 0.622$_{\pm0.042}$ & 0.712$_{\pm0.066}$ \\
\midrule
\kw{GraphConv} & 0.854$_{\pm0.011}$ & 0.877$_{\pm0.036}$ & 0.772$_{\pm0.041}$ & 0.650$_{\pm0.025}$ & 0.593$_{\pm0.035}$ & 0.845$_{\pm0.051}$ \\
\kw{Weave} & 0.791$_{\pm0.008}$ & 0.837$_{\pm0.065}$ & 0.741$_{\pm0.044}$ & 0.678$_{\pm0.024}$ & 0.543$_{\pm0.034}$ & 0.823$_{\pm0.023}$ \\
\kw{SchNet} & 0.750$_{\pm0.033}$ & 0.847$_{\pm0.024}$ & 0.767$_{\pm0.025}$ & 0.679$_{\pm0.021}$ & 0.545$_{\pm0.038}$ & 0.717$_{\pm0.042}$ \\
\kw{MGCN} & 0.734$_{\pm0.030}$ & 0.850$_{\pm0.064}$ & 0.707$_{\pm0.016}$ & 0.663$_{\pm0.009}$ & 0.552$_{\pm0.018}$ & 0.634$_{\pm0.042}$ \\
\kw{N}-\kw{Gram} & \cellcolor{lightgray}0.876$_{\pm0.035}$ & 0.912$_{\pm0.013}$ & 0.769$_{\pm0.027}$ & $-$\footnotemark & 0.632$_{\pm0.005}$ & 0.855$_{\pm0.037}$ \\
\midrule
\kw{MPNN} & 0.815$_{\pm0.044}$ & 0.913$_{\pm0.041}$ & 0.808$_{\pm0.024}$ & 0.691$_{\pm0.013}$ & 0.595$_{\pm0.030}$ & 0.879$_{\pm0.054}$ \\
\kw{DMPNN} & 0.852$_{\pm0.053}$ & \cellcolor{lightgray}0.919$_{\pm0.030}$ & \cellcolor{lightgray}0.826$_{\pm0.023}$ & \cellcolor{lightgray}0.718$_{\pm0.011}$ & \cellcolor{lightgray}0.632$_{\pm0.023}$ & \cellcolor{lightgray}0.897$_{\pm0.040}$ \\
\midrule
\kw{Concat+Mean} & 0.842$_{\pm0.004}$ & 0.930$_{\pm0.002}$ & 0.816$_{\pm0.003}$ & 0.721$_{\pm0.001}$ & 0.621$_{\pm0.007}$ & 0.882$_{\pm0.008}$ \\
\kw{Concat+Attn} & 0.832$_{\pm0.007}$ & 0.931$_{\pm0.006}$ & 0.819$_{\pm0.003}$ & 0.728$_{\pm0.002}$ & 0.632$_{\pm0.008}$ & 0.913$_{\pm0.009}$ \\
\midrule
\rowcolor{papergreen}\DualMPNN & 0.863$_{\pm0.002}$ & \textbf{0.938$_{\pm0.003}$} & 0.833$_{\pm0.001}$ & 0.729$_{\pm0.006}$ & \textbf{0.644$_{\pm0.003}$} & \textbf{0.930$_{\pm0.003}$} \\
\midrule
\rowcolor{papergreen}\DualMPNNplus & \textbf{0.892$_{\pm0.011}$} & 0.933$_{\pm0.006}$ & \textbf{0.836$_{\pm0.006}$} & \textbf{0.744$_{\pm0.005}$} & 0.639$_{\pm0.012}$ & 0.923$_{\pm0.007}$ \\
\bottomrule
\end{tabular}
}
\vskip -0.1 in
\label{baseline}
\end{table}
\footnotetext[4]{result not presented since N-Gram requires task-based preprocessing, which cannot stop in 10 days.}

\subsection{Ablation Studies on Key Design Choices}

\begin{figure}[!htb]
\begin{minipage}{0.48\textwidth}
  \centering
  \captionsetup{type=table}
  \caption{\footnotesize Ablation study on the variants of \DualMPNN.}
  \renewcommand\arraystretch{1.2}
  \resizebox{0.95\textwidth}{!}{
    \begin{tabular}{cccc}
    \toprule
     & \texttt{ToxCast} & \texttt{SIDER} & \texttt{ClinTox} \\
    \midrule
    No All & 0.718  & 0.644  & 0.852  \\
    Only Attention  & 0.728  & 0.646  & 0.901  \\
    Only Disagreement Loss & 0.722  & 0.648  & 0.863  \\
    \midrule
    \rowcolor{papergreen}\DualMPNN & \textbf{0.731}  & \textbf{0.652}  & \textbf{0.907}  \\
    \midrule
    \rowcolor{papergreen}\DualMPNNplus & \textbf{0.744}  & \textbf{0.639}  & \textbf{0.923} \\
    \bottomrule
    \label{tab:ablationstudy}
    \end{tabular}
}
\vspace{-1ex}
\end{minipage}
\hspace{0.05\textwidth}
\begin{minipage}{0.48\textwidth}
\centering
\includegraphics[width=0.97\textwidth]{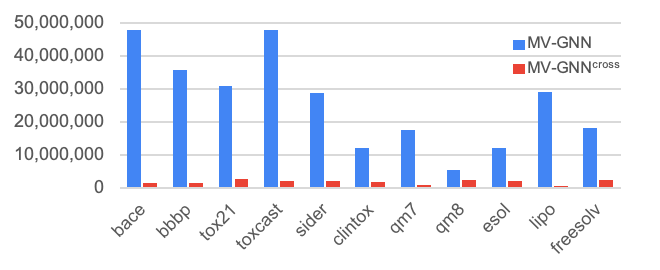}
\caption{\footnotesize Model parameters comparison.}
\vspace{-1ex}
\label{params}
\end{minipage}
\vspace{-1ex}
\end{figure}

This section  focuses on  the impacts of three key components in the proposed \DualMPNN models: the disagreement loss,  the shared self-attentive readout and the cross-dependent message passing scheme.
We report the results of three datasets with fixed train/valid/test sets to evaluate the impacts in Table~\ref{tab:ablationstudy}, which demonstrates the proposed multi-view  models overall performs the best on all three datasets.
Moreover, we find that both attention and disagreement loss can boost the performance compared with ``No All'' method. Particularly, when the self-attention mechanism is employed, the performance has already surpassed all the baseline models including \kw{MPNN} and \kw{DMPNN}, which proves that the molecular property is affected by the various atoms differently. Hence, the weights of atoms should not be considered equivalently. Overall, the proposed \DualMPNN models that adopts both disagreement loss and self-attention outperforms the other variants, indicating that the combination of them would significantly facilitate the model training.

\textbf{Effect of cross-dependent message passing.}  We plot the number of parameters in \DualMPNN and \DualMPNNplus in \cref{params}. 
It clearly indicates that \DualMPNNplus, while enjoying competitive performance,  needs much less amount of parameters than \DualMPNNplus.
Specifically, the average number of parameters of \DualMPNN is \emph{15.26} times of that of \DualMPNNplus. This  confirms that the  cross-dependent message passing scheme can significantly improve the expressive power of the model, by enabling a more efficient information communication scheme in the multi-view architecture.

\subsection{Case Study: Visualization of Interpretability Results}

\begin{figure*}[!h]
    \centering
    \includegraphics[width=0.23\textwidth]{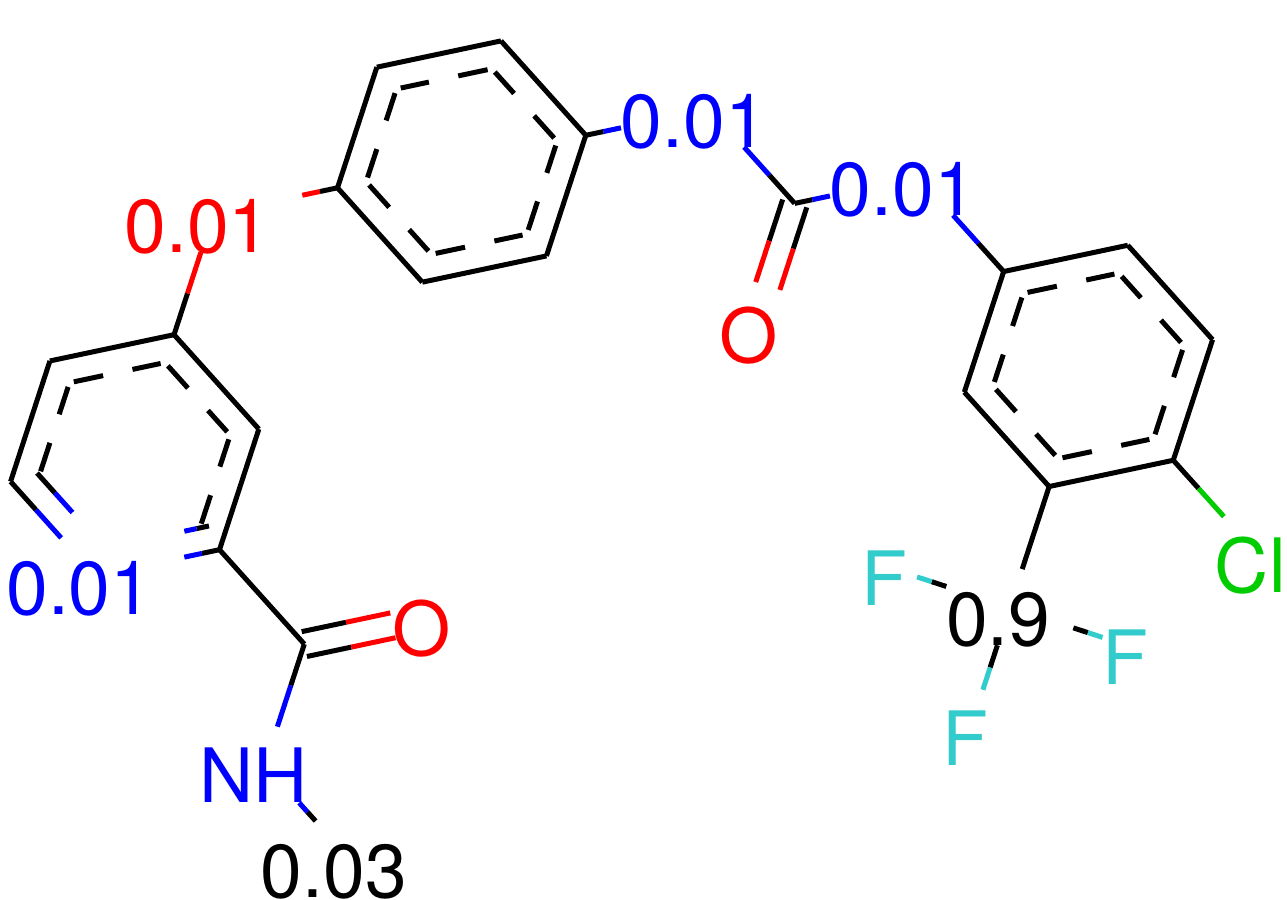}
    \includegraphics[width=0.23\textwidth]{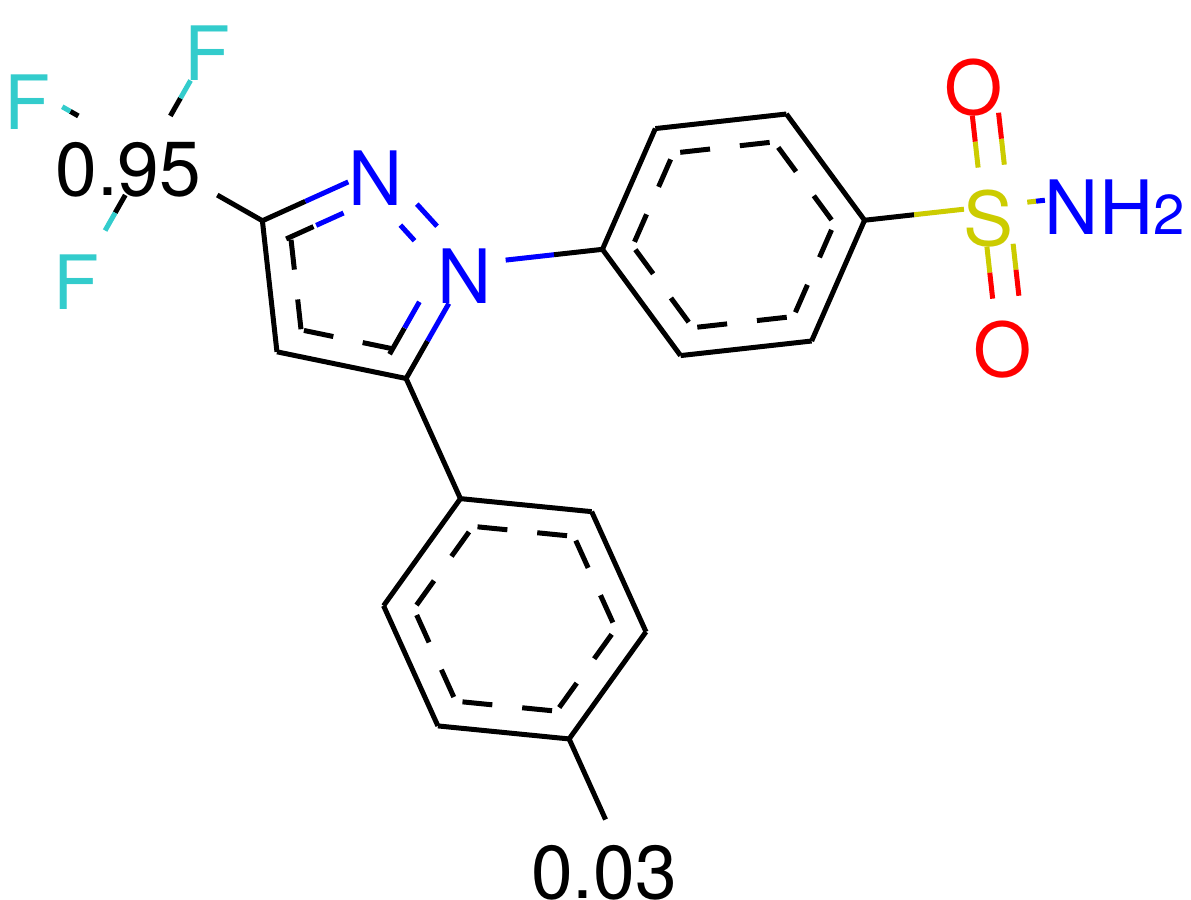}
    \includegraphics[width=0.23\textwidth]{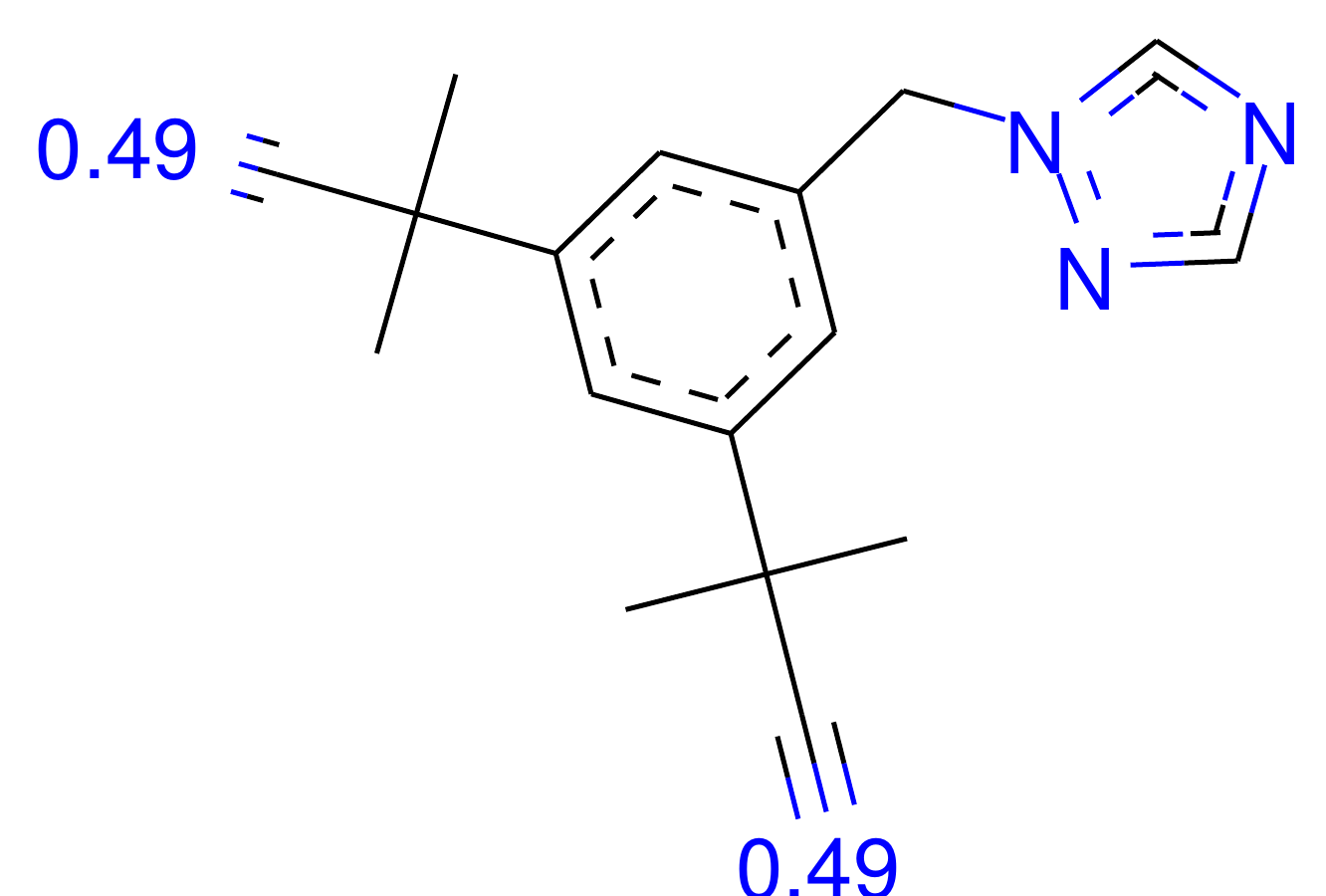}
    \includegraphics[width=0.23\textwidth]{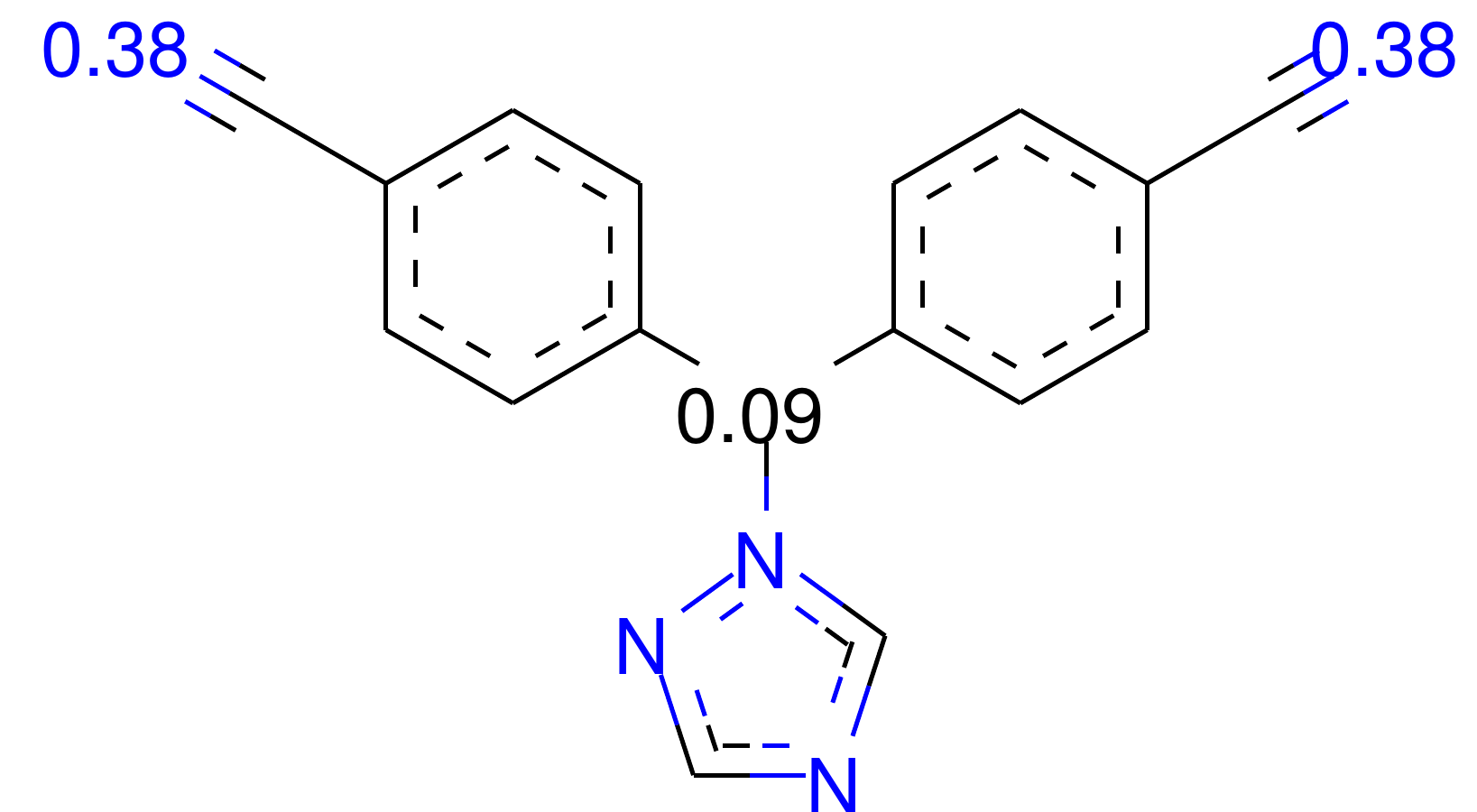}
    \vskip 0.1 in
    \caption{\footnotesize Visualization of attention values on  ClinTox data. Attention value smaller than 0.01 is omitted. Different color indicates different elements: black: C, \textcolor{blue}{blue: N}, \textcolor{red}{red: O}, \textcolor{ForestGreen}{green: Cl}, \textcolor{Peach}{yellow: S}, \textcolor{TealBlue}{sky-blue: F}.  First two molecules: the molecules with trifluoromethyl. Last two molecules: the molecules with cyanide.}
    \vskip -0.08 in
    \label{fig:attention_visual}
\end{figure*}

To illustrate the interpretability power of \DualMPNN, we visualize certain molecules with the learned attention weights of \DualMPNN associated with each atom within one molecule from the \texttt{Clintox} dataset, with toxicity as the labels.  \cref{fig:attention_visual} instantiates the graph structures of the molecules along with the corresponding atom attentions. The attention values lower than 0.01 are omitted. We observe that different atoms indeed react distinctively: 1) Most  carbon (C) atoms that are responsible for constructing the molecule topology have got zero attention value. It is because these kinds of sub-structures usually do not affect the toxicity of a compound. 2) Beyond that, \DualMPNN promotes the learning of the functional groups with impression on molecular toxicity, e.g., toxic functional group \emph{trifluoromethyl} and \emph{cyanide} are known responsible for the toxicity \cite{saarikoski1981influence}, which reveal extremely high attention value in \cref{fig:attention_visual}. These high attention values can be used to explain the toxicity of the molecules. Compared with the previous models, \DualMPNN is able to provide reasonable interpretability results for the predictions, which is crucial for the real drug discovery. 

\begin{wrapfigure}{r}{0.37\textwidth}
    \centering
    \includegraphics[width=0.37\textwidth]{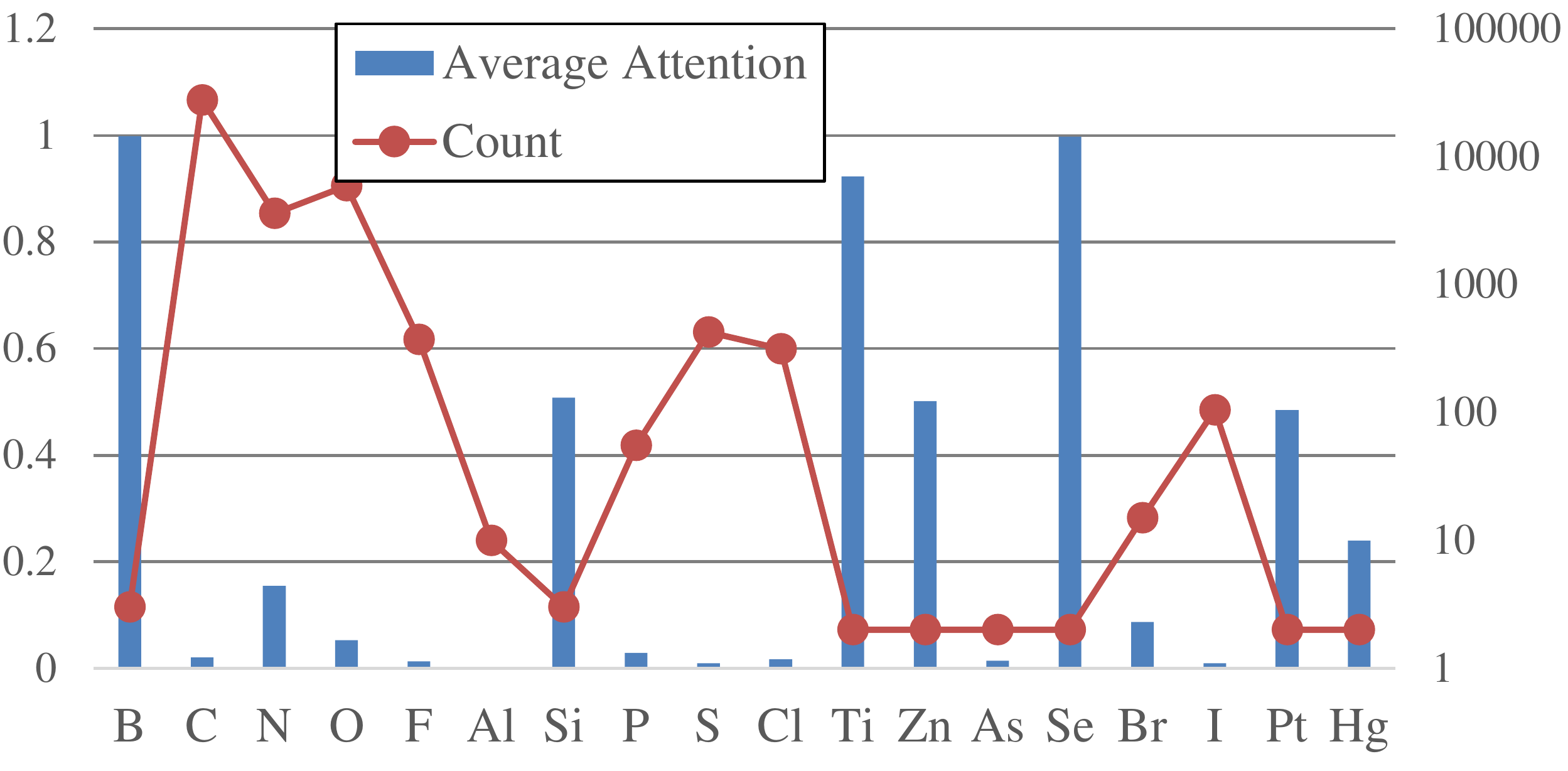}
    \caption{ \footnotesize Statistics of attentions in \texttt{ClinTox}. Left axis: the average attention value of the element. Right axis: the count of the element.}
    
    \label{fig:attention_statstics}
\end{wrapfigure}
Furthermore, we provide a comprehensive statistics of the attention values over the entire \texttt{ClinTox} dataset. Figure~\ref{fig:attention_statstics} demonstrates the average attention values and the total occurrences of each element. It is notable that, {1)} atoms with high frequency do not receive high attention. For example, atom C is an essential element to maintain the molecular topology, yet it does not have significant impact on the toxicity. {2)} atoms with low frequency but high attention values are generally heavy elements. For example, \kw{Hg} (Mercury) is widely known by its toxicity. The accompanied attention value of \kw{Hg} is relevantly high because it usually affects the toxic property greatly. 
Overall, the case study shows that the 
proposed \DualMPNN models are able to provide reasonable interpretability for the prediction results.


\section{Brief Related Work}
\label{sec_related_work}

Molecular representation learning and GNNs are extensively studied, which renders it very difficult comprehensively surveying all previous work. Here we only summarize some of the most related ones, and leave details in \cref{related_work}.
How to get accurate molecular representation is vital for molecular property prediction. Traditionally, chemical experts design a so-called {molecular fingerprint} manually based on their domain knowledge, e.g., ECFP \cite{rogers2010extended}. Several studies have exploited the deep learning approaches to improve the molecular representation.  One perspective is to take advantage of the molecular SMILES representation \cite{weininger1989smiles}. Based on the SMILES, \cite{Xu2017Seq2seqFA,jastrzkebski2016learning} apply RNN-based models to generate the molecular fingerprint. 
Another promising perspective is to explore the graph structure of a molecule by
graph neural networks (GNNs), which has attracted a surge of interest recently 
\cite{scarselli2008graph,kearnes2016molecular,schutt2017quantum,schutt2017schnet, liu2019n,xu2018powerful,ryu2018deeply,xiong2019pushing,velivckovic2017graph,gilmer2017neural,yang2019analyzing,klicpera_dimenet_2020,lu2019molecular}. 
In this line of work various GNN-based models have been proposed for generating molecular representations \citep{duvenaud2015convolutional,altae2017low,schutt2017quantum,schutt2017schnet,liu2019n}, e.g., \cite{duvenaud2015convolutional} applies convolutional networks on the molecular graphs to generate molecular fingerprint.

\vspace{-0.5em}
\section{Conclusions}

We propose  multi-view graph neural networks (\DualMPNN and \DualMPNNplus) for molecular property prediction.
Unlike previous attempts focusing exclusively on either atom-oriented graph structures or bond-oriented graph structures, our method, inspired by multi-view learning, takes both atom and bond information into consideration. 
We develop several techniques for the multi-view architecture: a shared self-attentive attention scheme enabling the interpretability power; a disagreement loss to restrain the distance between the outputs of the two views; a cross-dependent message passing scheme to enhance information communication between the views.  
Extensive experiments against SOTA models demonstrate that \DualMPNN and \DualMPNNplus outperform all baselines significantly, as well as equip with strong robustness. 
\bibliography{sample-base}

\newpage
\appendix

\section{Related Work in Details}
\label{related_work}

The most crucial part of addressing molecular property prediction problem is to get an accurate vector representation of the molecules. Relevant studies can be categorized into three aspects: hand-crafted molecular fingerprints based methods, SMILES sequence based techniques, and graph structure based techniques.

\textbf{Hand-crafted molecular fingerprints based methods.} The traditional feature extraction method enlists experts to design  molecular fingerprints manually, based on biological experiments and chemical knowledge \citep{morgan1965generation}, such as the property of molecular sub-structures. These types of fingerprint methods generally work well for particular tasks but lack universality. One representative approach is called circular fingerprints \citep{glen2006circular}. Circular fingerprints employ a fixed hash function to extract each layer's features of a molecule based on the concatenated features of the neighborhood in the previous layer. Extended-Connectivity Fingerprint (ECFP) \citep{rogers2010extended} is one of the most famous examples of hash-based fingerprints. The generated fingerprint representations usually go through machine learning models to perform further predictions, such as Logistic Regression \citep{kleinbaum2002logistic}, Random Forest \citep{breiman2001random}, and Influence Relevance Voting (IRV) \citep{swamidass2009influence}. Nonetheless, this type of hand-crafted fingerprint has a notable problem: since the characteristic of the hash function is non-invertible, it might not be able to catch enough information when being converted.

\textbf{SMILES sequence based techniques.} SMILES sequence based models, such as Seq2seq Fingerprint \cite{Xu2017Seq2seqFA}, spot the potentially useful information of the molecular SMILES sequence data by adequately training them using Recurrent Neural Networks (RNNs), in order to obtain the vector representation of the molecule. These vectors then go through other supervised models to perform property prediction, e.g., GradientBoost \citep{friedman2001greedy}. The SMILES-based models are inspired by the sequence learning in Natural Language Processing \citep{bordes2012joint}, which takes an unlabeled dataset as the input to convert a SMILES to a fingerprint, then recovers the fingerprint back to a sequence representation for better learning. 

\textbf{Graph structure based techniques.}
A molecule could be represented as a graph based on its chemical structure, e.g., consider the atoms as the nodes, and the chemical bonds between the atoms as the edges. Thus, many graph theoretic algorithms could be applied to represent a molecule by embedding the graph features into a continuous vector \citep{wu2018moleculenet,li2018adaptive,shang2018edge}. A noted study proposed the idea of neural fingerprints, which applies convolutional neural networks on graphs directly \citep{duvenaud2015convolutional}. The difference between neural fingerprints and circular fingerprints is the replacement of the hash function. Neural fingerprints apply a non-linear activated densely connected layer to generate the fingerprints. These kind of deep Graph Convolutional Neural Networks are established by learning a function on the graph node features and the graph structure matrix representation \citep{fout2017protein,li2018adaptive}. Other graph-based models such as the Weave model have also been proposed \citep{kearnes2016molecular}. The Weave model is another graph-based convolutional model. The key difference between the Weave model and neural fingerprints \citep{duvenaud2015convolutional} is the updating procedure of the atom features. It combines all the atoms in a molecule with their matching pairs instead of the neighbors of the atoms. More relevant research that focus on exploiting the molecular graphs with graph convolutional network have been studied recently, e.g., \citep{schutt2017quantum,schutt2017schnet} have involved the 3D information of the molecules to help exploit the molecular graph structure. Other attempts such as \citep{ryu2018deeply,xiong2019pushing} turn to develop aggregation weights learning schemes based on the prior knowledge of Graph Attention Network \citep{velivckovic2017graph}. Moreover, \citep{gilmer2017neural} proposes a framework to implement message passing process between each atom to form a molecular representation. Inspired by this work, \citep{yang2019analyzing,klicpera_dimenet_2020} convert the passing process to bond-wise instead of atom-wise. \cite{lu2019molecular} introduces multilevel graph structures based on the interactions between atom-pairs.


\section{More Details on \DualMPNN Models}
\label{app_details_models}
\begin{figure}[!h]
\centering  
\subfigure[\footnotesize \NodeMPN message passing]{
\label{fig:mpnn}
\includegraphics[width=0.4\textwidth]{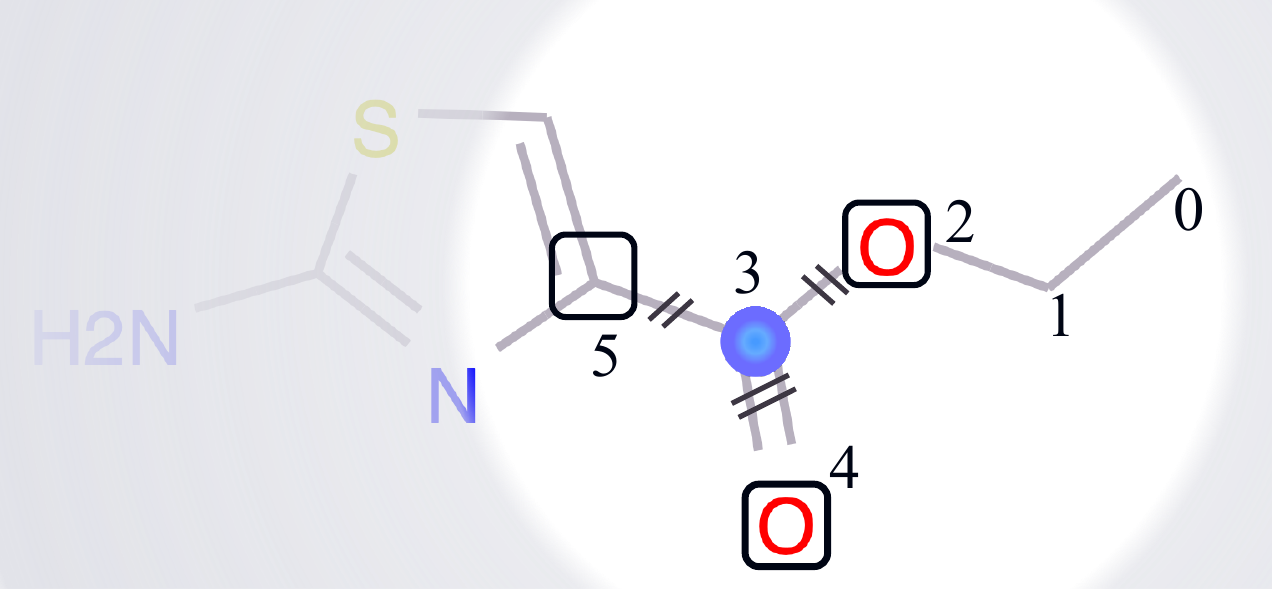}}
\hspace{0.1\textwidth}
\subfigure[\footnotesize \EdgeMPN message passing]{
\label{fig:dmpnn}
\includegraphics[width=0.4\textwidth]{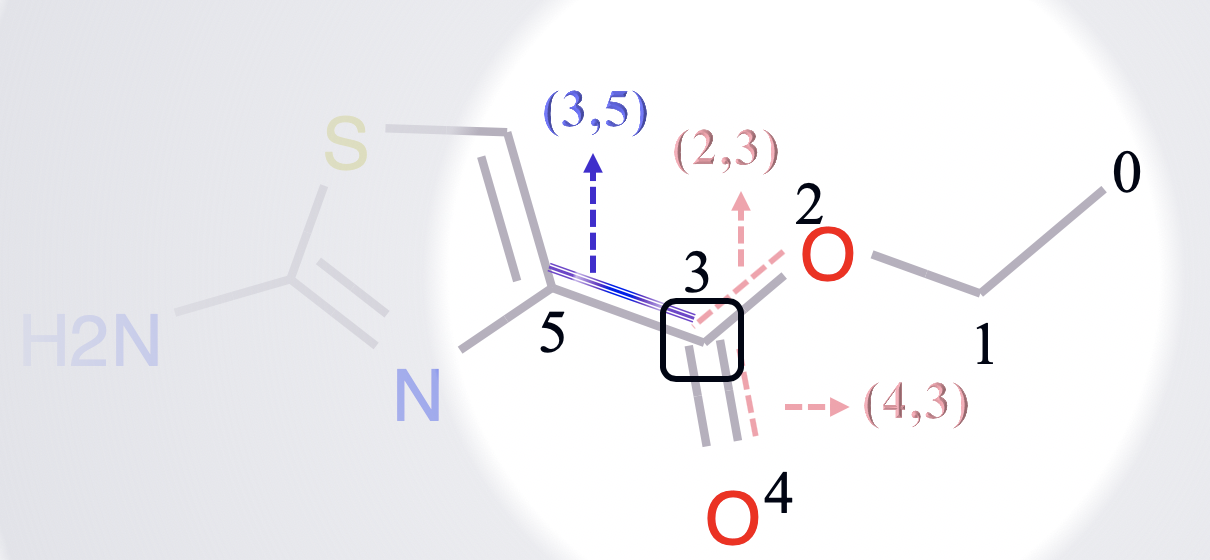}}
\caption{\footnotesize Examples of the message passing phase in \NodeMPN (\cref{fig:mpnn}) and \EdgeMPN (\cref{fig:dmpnn}).}
\label{Fig.message_passing}
\end{figure}

\DualMPNN models establish two sub-modules, \NodeMPN and \EdgeMPN. 
The process for each module can be categorized into three phases: neighbor aggregation, attached features collection, and message update. As shown in \cref{fig:mpnn}, for the \NodeMPN module, taking node $v_3$ as an example: 1) neighbor nodes aggregation: aggregating the node features of its neighbor nodes $v_2$, $v_4$ and $v_5$; 2)  getting the initial edge features as the attached features from the connected edge $\bm{e}_{23}$, ${\e}_{34}$, and $\e_{35}$; 3) updating the state of $v_3$ using \cref{eq_node_mpnn}. 

\cref{fig:dmpnn} demonstrates the edge message construction in the \EdgeMPN module. Take edge $\e_{35}$ as an example. 1) neighbor edges aggregation: aggregating the edge features of its neighbor edges, edge $\e_{23}$, edge $\e_{43}$; 2) getting the initial node information as the attached features from the endpoint node $v_2$ of edge $\e_{23}$, node $v_4$ of $\e_{43}$ ; 3) updating the message of $\e_{35}$ using \cref{equ_bondmpnn}.

\vskip -0.2 in

\section{Proof of Expressive Power }
\label{appendix_proof}

\begin{figure}[!h]
    \centering
    \includegraphics[width=0.46\textwidth]{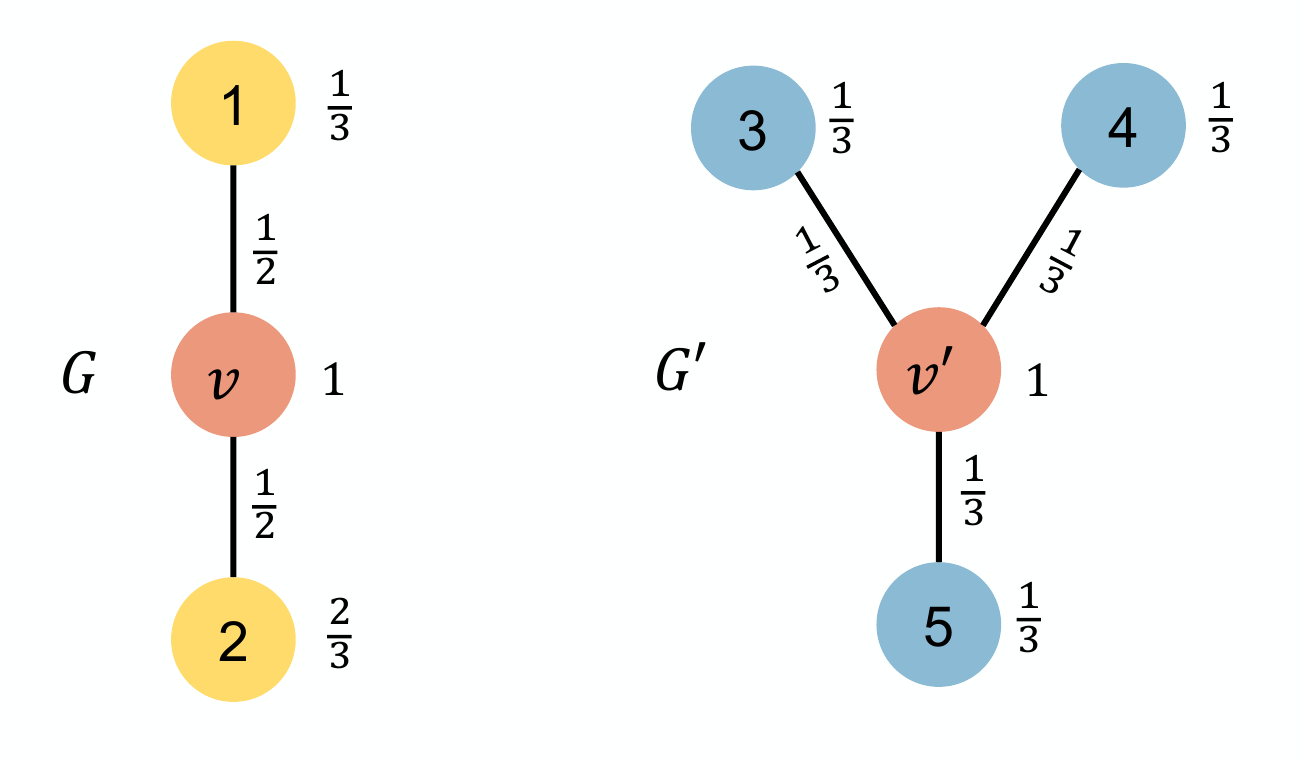}
    \vskip 0.05 in
    \caption{\footnotesize Example of the  two subgraph structures.} 
    \label{fig_counter_example}
    \vspace{-2ex}
\end{figure}

\begin{proof}[Proof of \cref{prop_expressiveness}]

Firstly we show that \DualMPNN is at least as powerful as  the Graph Isomorphism Network (GIN of \citet{xu2018powerful}). \DualMPNN involves both node message passing and edge message passing processes, which constitutes the two-view information flows. Suppose that one blocks the information flowing in the edge passing, say, by setting the initial hidden states of all edges to be $0$. At this moment, if one takes the the sum aggregation in \citet{xu2018powerful}  as the specific realization of the aggregation operation, then \DualMPNN recovers the GIN architecture. So we can conclude that \DualMPNN has at least the same expressive power as GIN. 

Then we prove that \DualMPNNplus  is strictly more powerful than  GIN. In order to illustrate this, we construct specific graph examples such that \emph{one} iteration of message passing in GIN cannot distinguish the nodes with different subgraph structures. However, \DualMPNNplus with the cross-dependent message passing scheme is able to discriminate the two nodes. 
To enable fair comparison, we assume that both GIN and \DualMPNNplus use the 
same aggregation function as in the GIN paper \citep{xu2018powerful}. That is, we use the sum aggregation with a parameter $\epsilon^\pare{l}$, i.e., $\agt^\pare{l}( \{ \h_v^\pare{l, k-1}, \h_u^\pare{l, k-1}, \e_{uv} |  u\in \mathcal{N}_v  \}) =  ( (1 + \epsilon^\pare{l}) \h_v^\pare{l, k-1} + \sum_{u\in \mathcal{N}_v} \h_u^\pare{l, k-1}) || (\sum_{u\in \mathcal{N}_v} \e_{uv})$ ($||$ is the concatenation operation). 

Assume there are two graphs $G, G'$ as shown in \cref{fig_counter_example}.
The two nodes therein, $v$ and $v'$ have different local subgraph structures. For simplicity, let all the initial node features, edge features and hidden states have dimensionality as 1. Specifically, for node features and hidden states, we have $\h_v = \x_v =1$, $\h_{v'} = \x_{v'} =1$, $\h_1 = \x_1 = \frac{1}{3}$,
$\h_2 = \x_2 = \frac{2}{3}$, $\h_3 =\h_4 =\h_5 = \x_3 =\x_4 =\x_5 =\frac{1}{3}$. For edge features and hidden states, one has $\h_{v1} = \h_{1v} = \e_{v1} = \frac{1}{2}$, $\h_{v2} = \h_{2v} = \e_{v2} = \frac{1}{2}$, $\h_{v'3} = \h_{3v'} = \e_{v'3} = \frac{1}{3}$, $\h_{v'4} = \h_{4v'} = \e_{v'4} = \frac{1}{3}$, 
$\h_{v'5} = \h_{5v'} = \e_{v'5} = \frac{1}{3}$. 

Under this setup, we can run one iteration of message passing by hand. Specifically,

--- For GIN, suppose its generalized version considers also the initial features. Then the message of node $v$ is $\m_v = (2+\epsilon, 1)$, where the first dimension indicates aggregated node hidden states, the second dimension indicates aggregated edge initial features. The message of node $v'$ is $\m_{v'} = (2+\epsilon, 1)$ as well. Since the state update function is \emph{injective}, after the state update, the hidden states of node $v$ and $v'$ will become the same, thus indistinguishable. 

--- For \DualMPNNplus, one complete iteration of message passing contains one edge message passing and one node message passing. Without loss of generality, let us take it as edge message passing followed by a node message passing. 

Consider edge message passing firstly. The edge messages for graph $G$ are: 
$\m_{v2} = (\overbrace{1/3}^{\text{node states sum}},  \overbrace{1 + \epsilon/2}^{\text{edge states sum} }, \overbrace{1/3}^{ \text{node features sum} }) $, 
$\m_{v1} = (\overbrace{2/3}^{\text{node states sum}},  \overbrace{1 + \epsilon/2}^{\text{edge states sum} }, \overbrace{2/3}^{ \text{node features sum} }) $. The two messages are different, the injective state update function will map them into different new states, suppose w.l.o.g. the new states are 
$\h_{v2} = 1, \h_{v2} = 2$. 

The edge messages for graph $G'$ are:  $\m_{v'3} =\m_{v'4} =\m_{v'5} = (\overbrace{2/3}^{\text{node states sum}},  \overbrace{1 + \epsilon/3}^{\text{edge states sum} }, \overbrace{2/3}^{ \text{node features sum} }) $. They will be mapped to the same new states, assume they are  $\h_{v'3} =\h_{v'4} =\h_{v'5} = 2/3$.  

Then consider node message passing with the newest edge hidden states. $\m_{v} = (\overbrace{2+\epsilon}^{\text{node states sum}},  \overbrace{3}^{\text{edge states sum} }, \overbrace{1}^{ \text{edge features sum} }) $, $\m_{v'} = (\overbrace{2+\epsilon}^{\text{node states sum}},  \overbrace{2}^{\text{edge states sum} }, \overbrace{1}^{ \text{edge features sum} })$. 

Now we have different messages for nodes $v$ and $v'$, so it will be mapped to different new hidden states by the injective  multi-layer perceptron (\mlp). 
Thus the two nodes become distinguishable under the cross-dependent message passing scheme of \DualMPNNplus. 

\end{proof}

\section{The Node/Edge Feature Extraction of the Molecules}
\label{feature_extraction}

The node/edge feature extraction contains two parts: 1) \textbf{node/edge messages}, which are constructed by aggregating neighboring nodes/edges features iteratively; 2) \textbf{molecule-level features}, which are the additional molecule-level features generated by RDKit to capture the global molecular information. It consists of 200 features for each molecule \cite{landrum2006rdkit}. Since we focus on the model architecture part, we follow the exact same protocol of \citep{yang2019analyzing} for the initial node (atom) and edge (bond) features selection, as well as the 200 RDKit features generation procedure. The atom features description and size are listed in \cref{tab:atomfea}, and the bond features are  documented in \cref{tab:bondfea}. The RDKit features are concatenated with the node/edge embedding, to go through the final MLP to make the predictions. 
\vskip -0.1 in

\begin{table}[!ht]
  \centering
  \renewcommand\arraystretch{1.2}
  \caption{\footnotesize Atom features \citep{yang2019analyzing}.}
  \vskip 0.1 in
  \resizebox{0.7\linewidth}{!}{
    \begin{tabular}{ccc}
    \toprule
          features & size & description \\
    \midrule
    atom type & 100  &  type of atom (e.g., C, N, O), by atomic number\\
    formal charge  & 5  & integer electronic charge assigned to atom  \\
    number of bonds & 6 & number of bonds the atom is involved in \\
    chirality & 4 & Unspeciﬁed, tetrahedral CW/CCW, or other.\\
    number of H & 5 & number of bonded hydrogen atoms\\
    atomic mass & 1 & mass of the atom, divided by 100\\
    aromaticity & 1 & whether this atom is part of an aromatic system\\
    hybridization & 5 &  sp, sp2, sp3, sp3d, or sp3d2\\
    \bottomrule
    \end{tabular}%
    }
  \label{tab:atomfea}%
\end{table}%

\begin{table}[!ht]
  \centering
  \renewcommand\arraystretch{1.2}
  \caption{\footnotesize Bond features \citep{yang2019analyzing}.}
  \vskip 0.1 in
  \resizebox{0.55\linewidth}{!}{
    \begin{tabular}{ccc}
    \toprule
          features & size & description \\
    \midrule
    bond type & 4  &  single, double, triple, or aromatic\\
    stereo & 6 &  none, any, E/Z or cis/trans \\
    in ring & 1 & whether the bond is part of a ring\\
    conjugated & 1 & whether the bond is conjugated\\
    \bottomrule
    \end{tabular}%
    }
  \label{tab:bondfea}%
\end{table}%


\section{Experimental Setup and Additional Results}

\subsection{Description of Dataset}
\label{dataset_description}

Table~\ref{datainfo} summaries the dataset statistics \citep{wu2018moleculenet}, including the property category, number of tasks and evaluation metrics of all datasets. Six datasets are used for classification, and five datasets  for regression. Noted, \texttt{ToxCast} contains 617 tasks, which makes it extremely time consuming to apply \kw{N}-\kw{Gram} model, since \kw{N}-\kw{Gram} requires task-based preprocess.
\vskip -0.1 in
\begin{table}[!h]
\caption{\footnotesize Datasets statstics. }
\vskip 0.1 in
\centering
\label{datainfo}
\renewcommand\arraystretch{1.2}
\resizebox{0.85\textwidth}{!}{
\begin{tabular}{cccccc}
\toprule
Category & Dataset & Task & \# Tasks & \# Graphs/Molecules & Metric \\
\midrule
Biophysics  & \texttt{BACE} & Classification & 1 & 1513 & AUC-ROC \\
\midrule
\multirow {5}{*}{Physiology} & \texttt{BBBP} & Classification& 1 & 2039 & AUC-ROC \\
 & \texttt{Tox21} & Classification & 12 & 7831 & AUC-ROC \\
 & \texttt{ToxCast} & Classification & 617 & 8576 & AUC-ROC \\
 & \texttt{SIDER} & Classification & 27 & 1427 & AUC-ROC \\
 & \texttt{ClinTox} & Classification & 2 & 1478 & AUC-ROC \\
\midrule
\multirow {2}{*}{\shortstack{Quantum\\Mechanics}} & \texttt{QM7} & Regression & 1 & 6830 & MAE \\
 & \texttt{QM8} & Regression & 12 & 21786 & MAE \\
\midrule
\multirow {3}{*}{\shortstack{Physical\\Chemistry}}& \texttt{ESOL} & Regression & 1 & 1128 & RMSE \\
 & \texttt{Lipophilicity} & Regression & 1 & 4200 & RMSE \\
 & \texttt{FreeSolv} & Regression & 1 & 642 & RMSE \\
\bottomrule
\end{tabular}
}
\end{table}
\vskip -0.15 in
\paragraph{Molecular Classification Datasets.}
\texttt{BACE} dataset is collected for recording compounds which could act as the inhibitors of human $\beta$-secretase 1 (BACE-1) in the past few years \cite{subramanian2016computational}. The Blood-brain barrier penetration (\texttt{BBBP}) dataset contains the records of whether a compound carries the permeability property of penetrating the blood-brain barrier \cite{martins2012bayesian}. \texttt{Tox21} and \texttt{ToxCast} \cite{richard2016toxcast} datasets include multiple toxicity labels over thousands of compounds by running high-throughput screening test on thousands of chemicals . \texttt{SIDER} documents marketed drug along with its adverse drug reactions, also known as the Side Effect Resource \cite{kuhn2015sider}. \texttt{ClinTox} dataset compares drugs approved through FDA and drugs eliminated due to the toxicity during clinical trials \cite{gayvert2016data}.

\paragraph{Molecular Regression Datasets.}
\texttt{QM7} dataset is a subset of GDB-13, which records the computed atomization energies of stable and synthetically accessible organic molecules, such as HOMO/LUMO, atomization energy, etc. It contains various molecular structures such as triple bonds, cycles, amide, epoxy, etc \cite{blum}. \texttt{QM8} dataset contains computer-generated quantum mechanical properties, e.g., electronic spectra and excited state energy of small molecules \cite{ramakrishnan2015electronic}. Both \texttt{QM7} and \texttt{QM8} contain 3D coordinates of the molecules along with the molecular SMILES. \texttt{ESOL} documents the solubility of compounds \cite{delaney2004esol}. \texttt{Lipophilicity} dataset is selected from ChEMBL database, which is an important property that affects the molecular membrane permeability and solubility. The data is obtained via octanol/water distribution coefficient experiments \cite{gaulton2011chembl}. \texttt{FreeSolv} dataset is selected from the Free Solvation Database, which contains the hydration free energy of small molecules in water from both experiments and alchemical free energy calculations \cite{mobley2014freesolv}.

\textbf{Dataset Splitting and Experimental Setting.} We apply the \emph{scaffold splitting} for all tasks on all datasets, which is more practical and challenging than 
random splitting. 
Random splitting is a common process to split the dataset into train, validation and test set randomly. 
However, it does not simulate the real-world scenarios for evaluating molecule-related machine learning methods \citep{bemis1996properties}. {Scaffold splitting} splits the molecules with distinct two-dimensional structural frameworks into different subsets \citep{bemis1996properties}, e.g. molecules with benzene ring would be split into one subset, which could be the train/validation/test set. This means that the validation/test dataset might contain molecules with unseen structures from the training dataset, which makes the learning much more difficult. Yet, it is  more  difficult for the learning algorithm to accomplish satisfactory performance, but from the chemistry perspective, it is more meaningful and consequential for molecular property prediction. To alleviate the effects of randomness and over-fitting, as well as to boost the robustness of the experiments, we apply cross-validation on all the experiments. All of our experiments run 10 randomly-seeded 8:1:1 data splits, which follows the same protocols of \cite{yang2019analyzing}.

\subsection{Baselines}
\label{baseline_model}


We thoroughly evaluate the performance of our methods with several popular baselines from both machine learning and chemistry communities. The post-fix $\kw{Reg}$ indicates the method for the regression task. Among them, Inﬂuence Relevance Voting (\kw{IRV}) is a K-Nearest Neighbor classifier, which assumes similar sub-structures reveal similar functionality \cite{swamidass2009influence}. \kw{LogReg}\cite{friedman2000additive} predicts the binary label by learning the coefficient combination of a logistic function based on the input features. Random Forest (\kw{RF}/\kw{RF\_Reg}) \cite{breiman2001random} is a decision tree based ensemble prediction model. The final result is generated by the ensemble of each decision tree prediction. \kw{GraphConv} \citep{duvenaud2015convolutional} is the vanilla graph convolutional model implementation by updating the atom features with its neighbor atoms' features. Compared with \kw{GraphConv}, \kw{Weave} \citep{kearnes2016molecular} model updates the atom features by constructing atom-pair with all other atoms, then combining the atom-pair features. \kw{SchNet} \citep{schutt2017schnet} and \kw{MGCN} \citep{lu2019molecular} explore the molecular structure by utilizing the physical information, the 3D coordinates of each atom. \kw{N}-\kw{Gram} \citep{liu2019n} proposes an unsupervised method to enhance the molecular representation learning by exploiting special attribute structure. \kw{MPNN} \citep{gilmer2017neural} and \kw{DMPNN} \citep{yang2019analyzing} perform the message passing scheme on atoms and bonds, respectively.

\subsection{Hyper-parameters}

We adopt Adam optimizer for model training. We use the Noam learning rate scheduler with two linear increase warm-up epochs and exponential decay afterwards \citep{vaswani2017attention}.

For \DualMPNN models on each dataset, we try 200 different hyper-parameter combinations via random search, and take the hyper-parameter set with the best test score. To remove randomness, we conduct  experiments 10 times with different seeds, along with the best-parameter set, to get the final result. The details of the hyper-parameters of the implementation of our models are introduced in Table \ref{tab:hyper}. 

\begin{table*}[htbp]
  \centering
  \renewcommand\arraystretch{1.3}
  \caption{\footnotesize Hyper-parameter Description.}
  \resizebox{\linewidth}{!}{
    \begin{tabular}{ccc}
    \toprule
          Hyper-parameter & Description&Range\\
    \midrule
    init\_lr &   initial learning rate of Adam optimizer and Noam learning rate scheduler & 0.0001\textasciitilde0.0004\\
    max\_lr  & maximum learning rate of Noam learning rate scheduler & 0.001\textasciitilde0.004\\
    final\_lr &  final learning rate of Noam learning rate scheduler & 0.0001\textasciitilde0.0004\\
    depth  &  number of the message passing hops ($K$) & 2\textasciitilde6\\
    hidden\_size  & number of the hidden dimensionality of the message passing network in two encoders ($d_{\text{hid}}$) & 7\textasciitilde19\\
    dropout & dropout rate & 0.5\\
    weight\_decay & weight decay percentage for Adam optimizer & 0.00000001\textasciitilde0.000001\\
    ffn\_num\_layers & number of the MLPs & 2\textasciitilde4\\
    ffn\_hidden\_size & number of the hidden dimensionality in the MLP & 7\textasciitilde19\\
    bond\_drop\_rate \cite{rong2020dropedge}& random remove certain percent of edges & 0\textasciitilde0.6\\
    attn\_hidden & number of hidden dimensionality in the self-attentive readout ($d_{\text{attn}}$) & 32\textasciitilde256\\
    attn\_out & number of output dimensionality in the self-attentive readout ($r$) & 1\textasciitilde8\\
    dist\_coff & the coefficient of the disagreement loss ($\lambda$) & 0.01\textasciitilde0.2\\
    \bottomrule
    \end{tabular}
    }
  \label{tab:hyper}%
\end{table*}%

\subsection{Model Size Comparison of \DualMPNN and \DualMPNNplus}
\label{model_parameters}

\cref{model_params} compares the parameter size of \DualMPNN and \DualMPNNplus for each dataset. As observed, \DualMPNNplus demands significantly less parameters. 
\vskip -0.1 in
\begin{table}[htbp]
\caption{\footnotesize Number of model parameters.}
\vskip 0.1 in
\centering
\label{model_params}
\renewcommand\arraystretch{1.0}
\resizebox{0.45\textwidth}{!}{
\begin{tabular}{ccc}
\toprule
Dataset & \DualMPNN & \DualMPNNplus \\
\midrule
\texttt{BACE} & 47,979,250 & 1,655,602 \\
\midrule
\texttt{BBBP} & 35,727,858 & 1,736,002 \\
\texttt{Tox21} & 31,057,826 & 2,757,024 \\
\texttt{ToxCast} & 47,877,458 & 2,358,034 \\
\texttt{SIDER} & 28,710,226 & 2,262,054 \\
\texttt{ClinTox} & 12,106,114 & 1,812,804 \\
\midrule
\texttt{QM7} & 17,628,514 & 1,029,602 \\
\texttt{QM8} & 5,493,314 & 2,418,424 \\
\midrule
\texttt{ESOL} & 12,106,114 & 2,248,802 \\
\texttt{Lipophilicity} & 29,069,026 & 826,402 \\
\texttt{FreeSolv}  & 18,266,626 & 2,457,602 \\
\bottomrule
\end{tabular}
}

\end{table}

\subsection{Additional Results of Classification Tasks}
\label{additional_classification}

\begin{figure}[htbp]
\centering  
\subfigure[\footnotesize Classification tasks.]{
\label{fig:cls_vis}
\includegraphics[width=0.45\textwidth]{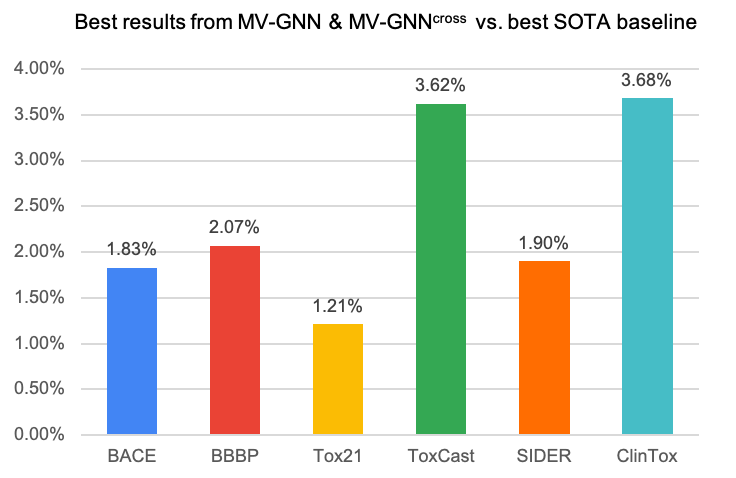}}
\hspace{0.05\textwidth}
\subfigure[\footnotesize Regression tasks.]{
\label{fig:reg_vis}
\includegraphics[width=0.45\textwidth]{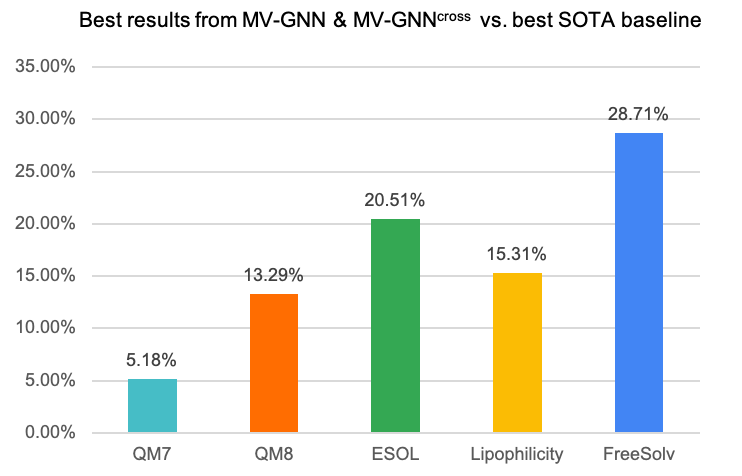}}
\caption{\footnotesize Improvement visualization between best \DualMPNN models with the best SOTA model on classification tasks (\cref{fig:cls_vis}) and regression tasks (\cref{fig:reg_vis}).}
\vspace{-2ex}
\label{fig:cls_reg_vis}
\end{figure}

\cref{fig:cls_vis} demonstrates the improvement between our models and the best SOTA model on the classification tasks, according to \cref{baseline}. As observed, \DualMPNN models are able to achieve up to 3.68\% on the \texttt{ClinTox} dataset.



\subsection{Additional Results of Regression Tasks}
\label{additional_regression}

Table~\ref{baseline_reg} reports the  results of \DualMPNN and \DualMPNNplus on regression tasks over 5 benchmark datasets and 7 baseline models. As we can see, \DualMPNN models achieve the best performance on regression tasks too. Recent graph studies on molecules generally focus on certain areas which lack universality. For example, \kw{SchNet} and \kw{MGCN} perform good on quantum mechanics datasets (\texttt{QM7} and \texttt{QM8}) since they utilizes the distances between atoms using the 3D coordinate information, but cannot capture sufficient molecule-level information to generate accurate molecular representations. On the other hand, our \DualMPNN models consistently achieve remarkable performance over all datasets. Specifically, our methods relatively improve  \textbf{28.7\%} over other models on the \texttt{FreeSolv} dataset, yet again, reveals the superiority and robustness of the multi-view architecture. The results of \kw{Concat+Mean} and \kw{Concat+Attn} on regression datasets also prove the effectiveness of \DualMPNN.


\cref{fig:reg_vis} illustrates the relative improvement from our model with other SOTAs, according to Table \ref{baseline_reg}. As shown in Figure~\ref{fig:reg_vis}, \DualMPNN models achieve average \textbf{16.6\%} improvement on the five regression benchmark datasets.

\begin{table}[htbp]
\centering
\captionsetup{type=table}
\renewcommand\arraystretch{1.2}
\caption{\footnotesize Performance comparison on regression tasks based on scaffold split (smaller is better). Best score is marked as \textbf{bold}, and the best baseline  is marked in gray background. Green cells indicate the results of our methods.}
\vspace{1.5ex}
\resizebox{0.88\textwidth}{!}{
\begin{tabular}{cccccc}
\toprule
Method & QM7 & QM8 & ESOL & Lipo & FreeSolv \\ 
\midrule
\kw{GraphConv} & 118.875 $_{\pm20.219}$ & 0.021 $_{\pm 0.001}$ & 1.068 $_{\pm 0.050}$ & 0.712 $_{\pm 0.049}$ & 2.900 $_{\pm 0.135}$ \\
\kw{Weave} & 94.688 $_{\pm 2.705}$ & 0.022 $_{\pm 0.001}$ & 1.158 $_{\pm 0.055}$ & 0.813 $_{\pm 0.042}$ & 2.398 $_{\pm 0.250}$ \\
\kw{SchNet} & \cellcolor{lightgray}74.204$_{\pm4.983}$ & 0.020$_{\pm0.002}$ & 1.045$_{\pm0.064}$ & 0.909$_{\pm0.098}$ & 3.215$_{\pm0.755}$ \\
\kw{MGCN} & 77.623$_{\pm4.734}$ & 0.022$_{\pm0.002}$ & 1.266$_{\pm0.147}$ & 1.113$_{\pm0.041}$ & 3.349$_{\pm0.097}$ \\
\kw{N}-\kw{Gram} & 125.630$_{\pm1.480}$ & 0.032$_{\pm0.003}$ & 1.100$_{\pm0.160}$ & 0.876$_{\pm0.033}$ & 2.512$_{\pm0.190}$ \\
\midrule
\kw{MPNN} & 112.960 $_{\pm 17.211}$ & 0.015 $_{\pm 0.002}$ & 1.167 $_{\pm 0.430}$ & 0.672 $_{\pm 0.051}$ & 2.185 $_{\pm 0.952}$ \\
\kw{DMPNN} & 105.775 $_{\pm 13.202}$ & \cellcolor{lightgray}0.0143 $_{\pm 0.0023}$ & \cellcolor{lightgray}0.980 $_{\pm 0.258}$ & \cellcolor{lightgray}0.653 $_{\pm 0.046}$ & \cellcolor{lightgray}2.177 $_{\pm 0.914}$ \\
\midrule
    \kw{Concat+Mean} & 72.532 $\pm$ 2.657 & 0.0129 $\pm$ 0.0005 & 0.806$\pm$0.040 & 0.610$\pm$0.024 & 2.003$\pm$0.317 \\
    \kw{Concat+Attn} & 73.132$\pm$3.845 & 0.0128$\pm$0.0005 & 0.809$\pm$0.043 & 0.601$\pm$0.015 & 2.026 $\pm$ 0.227 \\
\midrule
\rowcolor{papergreen}\DualMPNN & 71.325 \textbf{$_{\pm 2.843}$} & 0.0127 \textbf{$_{\pm 0.0005}$} & 0.8049 \textbf{$_{\pm 0.036}$} & 0.599 \textbf{$_{\pm 0.016}$} & 1.840 \textbf{$_{\pm 0.194}$} \\
\midrule
\rowcolor{papergreen}\DualMPNNplus & \textbf{70.358 $_{\pm 5.962}$} & \textbf{0.0124 $_{\pm 0.001}$} & \textbf{0.779$_{\pm 0.026}$} & \textbf{0.553 $_{\pm 0.013}$} & \textbf{1.552 $_{\pm 0.123}$} \\
\bottomrule
\end{tabular}
}
\vspace{-1ex}
\label{baseline_reg}
\end{table}

\end{document}